\documentclass[11pt]{article}
\usepackage{amsmath,amssymb}

\usepackage{epsfig,rotating,pifont}
\usepackage{graphicx}
\usepackage{caption}
\usepackage{subcaption}
\usepackage{dsfont}
\usepackage{tikz}
\usetikzlibrary{decorations.markings}
\usetikzlibrary{arrows}

\usepackage{hyperref}
\usepackage{cite}

\def\beq{\begin{eqnarray}}
\def\eeq{\end{eqnarray}}

\def\e{{\epsilon}}

\def\e{{\rm e}}

\newcommand{\bs}{\begin{split}}
\newcommand{\es}{\end{split}}

\newcommand{\Tr}{\,\mathrm{Tr}\,}            

\newcommand{\be}{\begin{equation}}
\newcommand{\ee}{\end{equation}}
\newcommand{\bea}{\begin{eqnarray}}
\newcommand{\eea}{\end{eqnarray}}
\newcommand{\bg}{\begin{gather}}

\newcommand{\bseq}{\begin{subequations}}
\newcommand{\eseq}{\end{subequations}}

\renewcommand{\ln}{\mathop{\rm ln}\nolimits}

\def\half{\frac{1}{2}}

\def\tr{\hbox{Tr}}

\def\be{\begin{eqnarray}}
\def\ee{\end{eqnarray}}
\def\lb{\label}

\def\e{\text{e}}

\begin{document}

\title{\textbf{Comparison theorems for causal diamonds}}

\vspace{2cm}
\author{ \textbf{
Cl\'ement Berthiere$^1$, Gary Gibbons$^{1,2,3}$  and  Sergey N. Solodukhin$^1$ }} 

\date{}
\maketitle
 \begin{center} \emph{ $^{1}$ Laboratoire de Math\'ematiques et Physique Th\'eorique  CNRS-UMR
7350 }\\
 \hspace{-0mm}
  \emph{F\'ed\'eration Denis Poisson, Universit\'e Fran\c cois-Rabelais Tours,  }\\
  \emph{Parc de Grandmont, 37200 Tours, France} 
 \hspace{-0mm}
 \end{center}
 \begin{center}
  \emph{$^{2}$ D.A.M.T.P., University of Cambridge, U.K.}
   \hspace{-0mm}
\end{center}
\begin{center}
  \hspace{-0mm}
  \emph{$^3$ LE STUDIUM, Loire Valley Institute for Advanced Studies,}\\
  \emph{ Tours and Orleans, France}
  \end{center}
{\vspace{-11cm}
\begin{flushright}
\end{flushright}
\vspace{11cm}
}



\begin{abstract}
\noindent { We formulate certain inequalities for  the geometric
quantities characterizing  causal diamonds in curved and Minkowski
spacetimes. These inequalities involve the red-shift factor which, 
as we show explicitly  in the spherically symmetric case,
is monotonic in the  radial direction and  it takes its  maximal
value at  the centre.  As a byproduct of our discussion we re-derive
Bishop's inequality without assuming the positivity of the spatial
Ricci tensor.  We then generalize our considerations to arbitrary,
static and not necessarily spherically symmetric, asymptotically flat
spacetimes.  In the case of spacetimes with a horizon our generalization
involves the so-called {\it domain of dependence}. The respective
volume, expressed in terms of the  duration measured by a distant observer
compared with the  volume of the domain in  Minkowski
spacetime, exhibits  behaviours which differ if  $d=4$ or  $d>4$. This
peculiarity of four dimensions  is due to the logarithmic subleading
term in the asymptotic expansion  of the metric near infinity. In
terms of the invariant duration measured by a co-moving observer
associated with the diamond we establish an inequality which is
universal for all $d$.  We suggest some possible applications of our
results including the comparison theorems for entanglement entropy,
causal set theory and fundamental limits on computation.

}
\end{abstract}

\vskip 2 cm
\noindent
\rule{7.7 cm}{.5 pt}\\
\noindent 
\noindent
\noindent ~~~ {\footnotesize e-mail:  clement.berthiere@lmpt.univ-tours.fr, gwg1@damtp.cam.ac.uk,  Sergey.Solodukhin@lmpt.univ-tours.fr}

\newpage
\setcounter{tocdepth}{2}
    \tableofcontents
\pagebreak

\newpage

\section{Introduction} \setcounter{equation}0 A causal diamond or
Alexandrov open set  is determined by the initial and final events $p$
and $q$ with respective proper-time separation $\tau$ as  a subset of
a Lorentzian spacetime $\{M^d, g\}$ of the form  \be I^+(p) \cap
I^-(q)\, , \lb{1}  \ee  where $I^+,I^- $ denotes chronological future
and past, respectively. Since the light rays propagation is the same
for conformally related metrics the causal diamond    depends only on
the conformal class of the Lorentzian metric $g$. However,  the
geometric quantities which characterize the diamond, such as its
$d$-volume  \be  V(p,q) = {\rm Vol}( I^+(p) \cap I^-(q     )=  \int
_{I^+(p) \cap I^-(q) } \sqrt{-g}\, d^d\,x \, ,  \lb{2} \ee depends
upon the metric itself.

The other important geometric quantity is 
the area $A(p,q)$ of the intersection $\dot I^+(p) \cap \dot
I^-(q) $ of the future light cone of $p$, ${\dot I}^+(q)=\partial
I^+(p)$ with the past light cone of $q$, ${\dot I}^+(q)=\partial I^+(p)$. 

Yet another quantity is the spatial volume defined for a hypersurface
having the intersection  $\dot I^+(p) \cap \dot I^-(q) $ of the future
light cone of $p$ with the past light cone of $q$ as their
boundary. There are many such hypersurfaces.   Among them,  there may
exist the one   with maximal volume ${\cal V}_{d-1}(p,q)$.  This
spatial volume is the third quantity  which characterizes the geometry
of the diamond. The earlier papers on the geometry of causal diamonds include
\cite{Myrheim:1978ce}, \cite{Gibbons:2007nm}, \cite{Gibbons:2007fd},
\cite{Solodukhin:2008qx}, \cite{papers}. The possible applications of these findings are the general theory of causal sets
as the foundation for the Quantum Gravity \cite{causalsets},  holography \cite{Freivogel:2014fqa}  and entanglement entropy \cite{entropy}.
A recent  application to geometry is considered in \cite{Park:2009ai}.

It should be noted that causal diamonds are important objects which
encode both the topological structure of the spacetime and its Riemann
geometry \cite{Hawking:1976fe}. As was shown in \cite{Gibbons:2007nm} one may use the
expansion of  the volume (\ref{2}) of a small causal diamond in powers
of $\tau$ and  reconstruct the Ricci curvature of spacetime. The area
$A(p,q)$ as well as the spatial volume ${\cal V}_{d-1}(p,q)$ have
similar expansions in powers of $\tau$ in terms of the Ricci tensor.

In this paper we are mostly interested in the case of infinitely large
diamonds, i.e. when the duration $\tau$ is taken to infinity.  Then
the diamond essentially becomes the whole spacetime and comparing its
volume  to that of a diamond of the same duration in Minkowski
spacetime would tell us whether the curved space has a bigger or
smaller volume compared to  flat space. Answering this question may
have various  interesting applications.  From the quantum mechanical
point of view the infinitely large diamond is the natural spacetime
domain in which the scattering of particles can be studied using the
concept of $S$-matrix. Thus the comparison problem which we study in
this paper may be useful for comparing the scattering amplitudes  in
Quantum Gravity in flat and in curved spacetimes.  

The curved spacetime may contain regions where the gravitational field
is strong. Since  the volume of a causal diamond is a non-local,
global, quantity,   the comparison of the volume with that of in
Minkowski spacetime may give us information about the strength of the
gravitational field.  As we shall see, this strength, in the cases
considered in the present paper, is essentially due to the central
red-shift.

Being interested in this sort of comparison problems for the
quantities $V_d(\tau)$, ${\cal V}_{d-1}(\tau)$  and $A_{d-2}(\tau)$ we
formulate certain inequalities  that relate these quantities with
their counter-parts in Minkowski spacetime. Some of these inequalities
are valid for a diamond of any size $\tau$ while for the others we
assume that the diamond is infinitely large.  Interestingly, the
universal inequalities are related to the monotonicity of certain
quantities constructed from the metric.  Most importantly, it is the
red-shift which in the spherically symmetric case is
monotonically decreasing along the radial direction
so that it takes its maximal value at the center of the spacetime.  In
many respects the comparison theorems which we formulate in this paper
are similar to those well known in the Euclidean Riemann geometry,
such as Bishop's inequality \cite{Bishop1}.

Some previously obtained results that are relevant to  the main subject of our paper should be mentioned.
In \cite{Gibbons:2007fd} it was observed  that the volume of a causal diamond in de Sitter spacetime (a solution to the Einstein equations with a positive cosmological constant), provided the duration of the diamond $\tau$ is fixed,
is a monotonic decreasing function of the 
cosmological constant $\Lambda$. So that the maximum of the volume corresponds to Minkowski spacetime and we have an inequality
\be
V_{dS}(\tau, \Lambda>0)< V_M(\tau,\Lambda=0)\, 
\lb{i1}
\ee
and similarly for the area
\be
A(\Lambda>0,\tau)< A(\Lambda=0,\tau)\, .
\lb{i11}
\ee
Then, in \cite{Solodukhin:2008qx} it was demonstrated that for a generic asymptotically de Sitter spacetime the volume $V(\tau,t_q)$ of a diamond of fixed duration $\tau$ is increasing function of 
cosmological time $t_q$. The asymptotic value of the volume is
that of in maximally symmetric de Sitter spacetime. Thus, asymptotically, for large values of cosmological time $t_q$, one has that
\be
V(\tau,t_q)\leq V_{dS}(\tau)\, , 
\lb{i2}
\ee
It was conjectured in \cite{Solodukhin:2008qx}  that this inequality is valid for any diamond in spacetime which is solution to vacuum Einstein equations with positive cosmological constant.

In an independent study \cite{ChoquetBruhat:2009fy} was formulated a light-cone theorem 
stating that the area of cross-sections of light-cones, in spacetimes satisfying suitable energy conditions, is smaller than or equal to that of the corresponding cross-sections in Minkowski, or de Sitter, or anti-de Sitter spacetime. Below we comment on a relation between our approach based on the geometry of causal diamonds and that of paper \cite{ChoquetBruhat:2009fy}.

This paper is organized as follows. In section 2 we set  up the geometric background of the problem and demonstrate the monotonicity of the red-shift
along the radial direction. In section 3 we formulate our main inequalities for the case when the spacetime is spherically symmetric. We also discuss the relation to the well-known in geometry Bishop's inequality and the relation to the light cone theorem of \cite{ChoquetBruhat:2009fy}. In section 4 we generalize this consideration to a static spacetime not assuming the spherical symmetry. In section 5 we further generalize our results to the case when the spacetime in question contains a black hole horizon. The possible applications are discussed in section 6. We conclude and summarize in section 7.

\section{Some preliminary geometry}
\setcounter{equation}0
\subsection{Metric and asymptotic conditions}
We start with the following spherically symmetric metric in $d$-dimensions
\begin{equation}
\label{noh}
ds^2=-f(r)\e^{2\gamma(r)}dt^2 + f(r)^{-1}dr^2 +r^2d\Omega^2_{d-2}\, , \qquad f(r)=1-\frac{2m(r)}{r^{d-3}}\, .
\end{equation}
The Einstein's equations $R_{\mu\nu}-\frac{1}{2}g_{\mu\nu}R=(d-2)\kappa_d T_{\mu\nu}$ produce equations for the functions $m(r)$ and $\gamma(r)$,
\begin{eqnarray}
\label{md}
\frac{d m(r)}{d r} & = & {\kappa_d}\,r^{d-2} T_{\hat{t}\hat{t}}\, , \\\label{gmd}
\frac{d \gamma(r)}{d r} & = & \kappa_d\,rf(r)^{-1}\left( T_{\hat{t}\hat{t}} + T_{\hat{r}\hat{r}}\right)  \, .
\end{eqnarray}
We shall assume some energy positivity conditions, 
\be
T_{\hat{t}\hat{t}}\geq 0 \, , \ \ T_{\hat{r}\hat{r}}\geq 0\, .
\lb{T}
\ee
This condition guarantees that the mass function 
\be
m(r)=\kappa_d\int_0^r dr' r'^{d-2}  T_{\hat{t}\hat{t}} \geq 0
\lb{m}\,,
\ee
so that the metric function $f(r)\le1$. Assuming also that the energy density $\rho=T_{\hat{t}\hat{t}}$ is finite at the origin, $r=0$, we find that $m(r)\sim r^{d-1}$ for small $r$
and hence $f(r=0)=1$. Together with the condition at spatial infinity,
\be
\lim_{r\rightarrow \infty} f(r)=1\,,
\lb{fr}
\ee
this fixes the boundary conditions for the function $f(r)$. We also assume that the spacetime does not contain black holes so that $f(r)$ never vanishes, $f(r)>0$.

On the other hand, the function $\gamma(r)$, as follows from equation (\ref{gmd}), is monotonic function of radial coordinate $r$,
\be
\frac{d \gamma(r)}{d r} \geq 0\, .
\lb{gamma}
\ee
The asymptotic boundary condition for $\gamma(r)$ is 
\be
\lim_{r\rightarrow\infty}\gamma(r)=0\,,
\lb{gr}
\ee
so that the metric component $g_{tt}=-f(r)e^{2\gamma(r)}$ approaches $-1$ at the spatial infinity. With this asymptotic condition the metric (\ref{noh}) describes
asymptotically flat spacetime.
The monotonicity of $\gamma(r)$ implies that  $\gamma(r)<0$ is negative and at the origin it takes a negative value $\gamma(r=0)=\gamma_0<0$.

\subsection{Monotonicity of the red-shift}
As we have seen the mass function $m(r)$ is monotonic in $r$. The metric function $f(r)$ on the other hand is not monotonic. Indeed, it takes value $1$ both at
$r=0$ and $r=\infty$. One can construct using $m(r)$ and $\gamma(r)$ another function which is monotonic in $r$. This function is the $(tt)$-component of the metric,
\be
-g_{tt}(r)=f(r)e^{2\gamma(r)}\, .
\lb{gtt}
\ee
The derivative of this component with respect to $r$ reads
\be
-\frac{d g_{tt}(r)}{d r}=\frac{2e^{2\gamma(r)}}{r^{d-2}}\left((d-3)m(r)+\kappa_d\, r^{d-1} T_{\hat{r}\hat{r}} \right)\geq 0\, .
\lb{dg}
\ee
With the energy conditions (\ref{T}) the right hand side of this relation is a non-negative function. Thus,  (\ref{gtt}) is monotonically increasing function of $r$.
Using the boundary conditions we have imposed on $f(r)$ and $\gamma(r)$ in the previous section we have the following inequalities
\be
   e^{2\gamma_0}\,  \leq f(r)e^{2\gamma(r)}\leq \, 1\, .
\lb{ineqf}
\ee
The value of $(tt)$-component of the metric at point $r$  has interpretation as a red-shift compared to an observer at infinity,
\be
\frac{1}{1+z}=\sqrt{-g_{tt}(r)}  \, .
\lb{z}
\ee
The value $\gamma_0$ then can be expressed in terms of the central red-shift, $z_c=z(r=0)$, as 
\be
1+z_c=e^{-\gamma_0}\, .
\lb{zc}
\ee
Eq. (\ref{dg}) then means that the red-shift is maximal at the center of the spacetime, $r=0$,
\be
0<z(r)\leq z_c\, .
\lb{zzc}
\ee

\section{Comparison theorems}
\setcounter{equation}0
\subsection{The causal diamond}
Let us first introduce the tortoise coordinate $y$,
\be
y=\int_0^r  e^{-\gamma(r')}\frac{dr'}{f(r')}\, .
\lb{y}
\ee
We then consider two events $p$ and $q$, both at the origin $r=0$, such that the invariant time interval between  them is $\tau$. The respective interval in time $t$ is
$T$, the relation between them is $\tau=e^{\gamma_0}T=1/(1+z_c) T$. The light cone $I^-(q)$ is defined by equation $t=T/2-y$ while the light cone $I^+(p)$ is defined by
$t=-T/2+y$.  On the plane $(t,y)$ they intersect at $y=T/2$. The respective value of radial coordinate $r(T/2)$ is found from equation
\be
T/2=\int_0^{r(T/2)}e^{-\gamma(r')}\frac{dr'}{f(r')}\, 
\lb{T}
\ee
or, equivalently, in terms of invariant time interval
\be
\tau/2=\int_0^{r(T/2)}e^{(\gamma_0-\gamma(r'))}\frac{dr'}{f(r')}\, .
\lb{tau1}
\ee
The causal diamond formed by the points $p$ and $q$ is  spherically symmetric.
The volume of this  diamond is
\be
V(\tau)=2 \Omega_{d-2}\int_0^{T/2}dt \int_0^{T/2-t} dy \, r^{d-2}(y) f(y)e^{2\gamma(y)} ,
\lb{V}
\ee
where $\Omega_{d-2}$ is the area of $(d-2)$-sphere of unite radius.
We also compute the area and the spatial volume of the diamond,
\be
A(\tau)=\Omega_{d-2} r^{d-2}(T/2)\, , \qquad {\cal V}(\tau)=\Omega_{d-2}\int_0^{r(T/2)}dr'\,\frac{r'^{d-2}}{\sqrt{f(r')}}\, .
\lb{AV}
\ee

We can use eqs.~(\ref{T}) and  (\ref{tau1}) and deduce some inequalities. Since $f(r)\le1$ and $e^{-\gamma(r)}\ge1$ we find that
\be
r(T/2)\leq T/2=e^{-\gamma_0}\, \tau/2 \, .
\lb{rT}
\ee
This inequality is universal valid for any value of $T$. The other inequality can be derived for large values of $T$. Indeed, in the limit of
large $T$ (respectively large $r(T/2)$) we have that $f(r)$ approaches $1$ and $\gamma(r)$ approaches $0$ so that we find from (\ref{tau1})
\be
\tau/2=e^{\gamma_0}r(T/2)\leq r(T/2)\, .
\lb{taur}
\ee
We stress once again that (\ref{taur}) is valid for large (infinite) values of $T$ (or $r(T/2)$).

\subsection{Inequalities for volume of causal diamond}
In order to arrive at our first inequality we differentiate equation (\ref{V}) twice with respect to $\tau$ (we remind the reader that $\tau=T e^{\gamma_0}$). We find
\be
\frac{d^2 V(\tau)}{d\tau^2}=\frac{\Omega_{d-2}}{2} \, r^{d-2}(T/2)f(T/2)e^{2(\gamma(T/2)-\gamma_0)}\, .
\lb{dV}
\ee
Using the inequality (\ref{ineqf}) we obtain
\be
\frac{\Omega_{d-2}}{2} \, r^{d-2}(T/2) \leq \frac{d^2 V(\tau)}{d\tau^2}\leq \frac{\Omega_{d-2}}{2} \, r^{d-2}(T/2)e^{-2\gamma_0}\, .
\lb{VV}
\ee
Both inequalities are universal and valid for any values of $T$. For the upper bound in (\ref{VV}) we use (\ref{rT}) and find
\be
\frac{d^2 V(\tau)}{d\tau^2}\leq \frac{\Omega_{d-2}}{2} \, \left(\frac{\tau}{2}\right)^{d-2} e^{-d\gamma_0}\, .
\lb{VVd}
\ee
Integrating this inequality twice with respect to $\tau$ and taking into account that $V(\tau=0)=0$ we find the desired inequality for the volume of causal diamond
\be
V(\tau)\leq V_{M}(\tau)e^{-d\gamma_0}=(1+z_c)^d\,V_M(\tau) \, ,
\lb{V<M}
\ee
where $V_M(\tau)$ is volume of causal diamond of duration $\tau$ in Minkowski spacetime. This inequality is universal, valid for any values of $\tau$.

Now consider the lower bound in (\ref{VV}) and use the inequality (\ref{taur}) we arrive at
\be
\frac{\Omega_{d-2}}{2} \, \left(\frac{\tau}{2}\right)^{d-2}\leq \frac{d^2 V(\tau)}{d\tau^2}\, .
\lb{tauV}
\ee
Integration over $\tau$ then gives us the following inequality
\be
V_M(\tau)\leq V(\tau)\, .
\lb{M<V}
\ee
Combining the two inequalities, (\ref{V<M}) and (\ref{M<V}), we find
\be
V_M(\tau) \leq V(\tau)\leq (1+z_c)^d\,V_M(\tau)\, .
\lb{MVM}
\ee
These inequalities compare the volume of a causal diamond in a curved spacetime with that of in Minkowski spacetime. Interestingly, the comparison involves the
central red-shift $z_c$.
We notice that the upper bound in (\ref{MVM}) is universal while the lower bound is valid only in the limit of infinite $\tau$.

\subsection{More inequalities:  the area and spatial volume }
Consider now the area $A(\tau)$ (\ref{AV}). Using inequalities (\ref{rT}) and (\ref{taur}) 
we find the following inequalities for the area
\be
A_M(\tau)\leq A(\tau)\leq (1+z_c)^{d-2}A_M(\tau) \, ,
\lb{AM}
\ee
where $A_M(\tau)=\Omega_{d-2}(\tau/2)^{d-2}$ is the area in the case of the diamond in  Minkowski spacetime.

In order to get the inequalities for the spatial volume $\cal{V}(\tau)$ we first re-write (\ref{AV}) in a slightly different form
\be
{\cal{V}}(\tau)=\Omega_{d-2}\int_0^{T/2}dy\, r^{d-2}(y)\sqrt{f(y)}e^{\gamma(y)}\, .
\lb{Vsp}
\ee
Then we use the red-shift inequality (\ref{ineqf}) and find
\be
\Omega_{d-2}\,e^{\gamma_0}\int_0^{T/2} dy\,r^{d-2}(y) \leq {\cal{V}}(\tau)\leq \Omega_{d-2}\int_0^{T/2}dy\, r^{d-2}(y)\, .
\lb{Vspn}
\ee
Since $y\ge r(y)$, as follows from (\ref{rT}), one finds
\be
\int_0^{T/2}dy\, r^{d-2}(y) < \int_0^{T/2}dy\,y^{d-2} = \frac{1}{d-1}\left(\frac{T}{2}\right)^{d-1}=e^{-(d-1)\gamma_0}\frac{1}{d-1}\left(\frac{\tau}{2}\right)^{d-1}\, .
\lb{rs}
\ee
On the other hand, we have
\be
\int_0^{T/2}dy\, r^{d-2}(y) >\int_0^{r(T/2)}dr'\, r'^{d-2}=\frac{1}{d-1}r^{d-1}(T/2)\geq \frac{1}{d-1}\left(\frac{\tau}{2}\right)^{d-1}\, ,
\lb{ls}
\ee
where in the last inequality we have used (\ref{taur}) valid for large $\tau$.

Finally, combining the inequalities (\ref{Vspn}), (\ref{rs}), and (\ref{ls}) we  arrive at
\be
{\cal{V}}_M(\tau)\leq {\cal{V}}(\tau)\leq  (1+z_c)^{d-1}\,\mathcal{V}_M(\tau) \, ,
\lb{spatV}
\ee
where ${\cal{V}}_M(\tau)=\frac{\Omega_{d-2}}{d-1}(\tau/2)^{d-1}$ is the spatial volume of a diamond of duration $\tau$ in Minkowski spacetime.
As before, the upper bound is universal while the lower one is valid for large $\tau$.
Equations (\ref{MVM}), (\ref{AM}), and (\ref{spatV}) are the comparison inequalities and our main result.

\subsection{Relation to Bishop's inequality}
\lb{Bishop}
The results of the previous section for the spatial volume may be interpreted in terms of Bishop's inequality. 
The cross-section of hypersurface of $t=0$ and the diamond is a ball of radius
\be
{\cal{R}}=\int_0^{r(T/2)} \frac{dr'}{\sqrt{f(r')}}\ge r(T/2)\, ,
\lb{B1}
\ee
where we use that $f(r)\leq 1$.
Volume of this ball is 
\be
{\cal{V}}({\cal{R}})=\Omega_{d-2}\int_0^{r(T/2)}dr' \frac{r'^{d-2}}{\sqrt{f(r')}}\leq \frac{\Omega_{d-2}}{d-1}r^{d-1}(T/2)\leq \frac{\Omega_{d-2}}{d-1} {\cal{R}}^{d-1}={\cal{V}}_M({\cal{R}})\, ,
\lb{B2}
\ee
where ${\cal{V}}_M({\cal{R}})$ is volume of  a ball of radius ${\cal{R}}$ in flat $(d-1)$-dimensional Euclidean space. 

Inequality (\ref{B2}) is just another form of the upper bound in 
(\ref{spatV}). On the other hand, (\ref{B2}) coincides with the well-known Bishop's inequality. Bishops' proof, and those that followed, 
assumed that  the Ricci tensor of the $(d-1)$-hypersurface is positive, $^{(d-1)}R_{ij}\geq K\gamma_{ij}$, where $K$ is a positive constant. We notice, however, that this condition does not hold in our case.  In dimension $d\geq 4$,
we have 
\be
^{(d-1)}R_{\hat r \hat r} = - \frac{d-2}{r^{d-1}} \left((d-3)m(r)-r \frac{dm(r)}{dr}\right) \,,
\quad ^{(d-1)}R_{\hat \theta_i \hat \theta_i}= \left((d-3)m(r)+ r \frac{dm(r)}{dr}\right)/r^{d-1} \, .
\lb{Ricci}
\ee
 Clearly, the components of the Ricci tensor (\ref{Ricci}) cannot be positive at the same time.
In fact, for the Schwarzschild metric  $^{(d-1)}R_{\hat r \hat r}<0$ while $^{(d-1)}R_{\hat \theta_i \hat \theta_i}>0$.
It is interesting that despite the fact that the assumptions made by 
 Bishop  do not hold, his  inequality still holds for the hypersurface $t=0$.
This example suggests that perhaps the conditions in the Bishop theorem could be relaxed and admit a Ricci tensor which is not sign definite.

In order to get a hint of a possible generalization of our result, 
let us re-derive it in a slightly different way not making use of any 
inequalities for
function $f(r)$. The area of surface at radius $r$ is $A(r)=\Omega_{d-2}r^{d-2}$. The invariant geometric radius of this sphere is $\rho=\int_0^rdr'/\sqrt{f(r')}$.
Let us define 
\be
\theta(r)=A^{-1}\frac{dA}{d\rho}=(d-2)r^{-1}\sqrt{f(r)}\, .
\lb{x}
\ee
Differentiating this equation once again with respect to $\rho$ we obtain
\be
\frac{d\theta}{d\rho}=-\frac{\theta^2}{d-2}+\frac{(d-2)}{2r}f'_r\, 
\lb{x2}
\ee
or, using (\ref{Ricci}), this can be re-written  as
\be
\frac{d\theta}{d\rho}=-\frac{\theta^2}{d-2}-\, ^{(d-1)}R_{\hat r \hat r}\, .
\lb{x3}
\ee
This is a Raychaudhuri type of equation. As we noticed above Ricci tensor $ ^{(d-1)}R_{\hat r \hat r}$ is negative, at least outside the matter source.
Representing $A(r)=\Omega_{d-2} w^{d-2}$ and hence $\theta=(d-2)w^{-1}w'_\rho$ this equation can be rewritten as equation for $w(\rho)$ 
\be
w''_\rho=-R_{\hat r \hat r}w\, .
\lb{x4}
\ee
Integrating this equation from $\rho$ to infinity and taking into account the asymptotic boundary condition $w(\rho)\rightarrow\rho$ if  $\rho\rightarrow \infty$ we find
\be
1-w'_\rho=-\int_\rho^\infty R_{\hat r \hat r}(\rho')w(\rho') d\rho'\, .
\lb{x5}
\ee
The right hand side of this equation is positive and thus we conclude that
\be
\frac{dw}{d\rho}<1\, .
\lb{x6}
\ee
Integrating this equation we obtain an inequality,
\be
w(\rho)<\rho
\lb{x7}
\ee
or, equivalently,
\be
A(\rho)< \Omega_{d-2}\rho^{d-2}=A_M(\rho)\, ,
\lb{x8}
\ee
where $A_M(\rho)$ is the area of $(d-2)$-sphere of radius $\rho$ in Minkowski spacetime. Integrating (\ref{x8}) with respect to $\rho$ and taking into account that the $(d-1)$-volume ${\cal{V}}({\cal R})=\int_0^{\cal{R}}d\rho A(\rho)$ we arrive at the inequality (\ref{B2}).

\subsection{Relation to light cone theorem}
In this section we make contact with the theorem proved in \cite{ChoquetBruhat:2009fy}.
Let $s$ be an affine parameter along 
a light cone normalised such that
\be
\frac{d x^\alpha}{ds} T^\beta g_{\alpha \beta}=- 1 \,,
\ee
where $T^\alpha$ is a unit timelike vector at the tip. That is $
g_{\alpha \beta} T^\alpha T^\beta = -1 $.
If the tip is at the center then
$
\displaystyle T^\alpha = \frac{1}{\sqrt{-g_{tt}}}\delta ^\alpha_t  
$
and hence
\be
\frac{dt}{ds} =\sqrt{-g_{tt}} = \frac{1}{1+z_c} = \sqrt{f(0)} \, e^{\gamma(0)} \, .   
\ee
Thus, we have the following inequalities
\be
s= (1+z_c) t = (1+z_c) y(s) \ge (1+z_c) r(s) \ge r(s) \, . 
\lb{g1}
\ee
From the metric, the area $A(s)$ of the cross section of the cone at the value $s$ of the affine parameter is
\be
A(s) = \Omega_{d-2} r^{d-2}(s)\, . 
\ee
Thus, using (\ref{g1}) we find that
\be
A(s) \le  \Omega_{d-2} s^{d-2} \,. 
\ee
This is a special case of the result proved in \cite{ChoquetBruhat:2009fy}. On the other hand, this inequality is just another form
of the upper bound in (\ref{AM}).

\subsection{Adding a cosmological constant}
Cosmological observations strongly suggest the presence of a cosmological constant $\Lambda$ and much theoretical research involves this constant. Therefore it is interesting to formulate comparison theorems for spacetimes which are asymptotically de Sitter and anti-de Sitter. Then comparisons will be made with the maximally symmetric de Sitter and anti-de Sitter spacetimes.

In de Sitter and anti-de Sitter spacetimes the causal diamond formed by the points $p$ and $q$, both at $r=0$, is spherically symmetric and the invariant time interval is $\tau = T$, just as in the Minkowski case. The metrics are the following,
\be
ds_{(a)dS}^2=-f(r)dt^2 + f(r)^{-1}dr^2 +r^2d\Omega^2_{d-2}\, , \qquad f(r)=1\mp r^2/\ell^2 \, ,
\ee
where $\ell^2 = \frac{(d-1)(d-2)}{2|\Lambda|}$ and the sign (plus)minus corresponds to (anti-) de Sitter. 

The volume of the causal diamond in $d$-dimensional de Sitter spacetime is (\ref{V})
\be
V_{dS}(\tau)=\frac{\Omega_{d-2}\ell^d}{d(d-1)}\left(2\tanh^{d}\frac{\tau}{2\ell}+d~B\left(\tanh^{2}\frac{\tau}{2\ell}; 1+d/2,0\right)   \right) \,,
\ee 
where $B(z;a,b)$ is the incomplete Beta function. The area (\ref{AV}) is given by
\be
A_{dS}(\tau)=\Omega_{d-2}\ell^{d-2} \tanh^{d-2}(\tau/2\ell)\,.
\ee
We do not give here the explicit formula of spatial volume (\ref{AV}) as it is somehow complicated (involving hypergeometric functions) and not very useful. For the causal diamond in anti-de Sitter spacetime, one finds the volume and area and spatial volume by making the change $\ell \rightarrow i\ell$. Some inequalities can already be formulated. For the de Sitter case we have
\be
f(r) = 1-r^2/\ell^2 \leq 1\,,
\ee
from which we deduce
\be
V_{dS}(\tau) \leq V_{M}(\tau)\,, \qquad  {\cal{V}}_{dS}(\tau) \leq {\cal{V}}_{M}(\tau)\,, \qquad  A_{dS}(\tau) \leq A_{M}(\tau)\,.
\ee
For the anti-de Sitter case we have
\be
f(r) = 1+r^2/\ell^2 \geq 1\,,
\ee
from which we find
\be
V_{adS}(\tau) \geq V_{M}(\tau)\,, \qquad  {\cal{V}}_{adS}(\tau) \geq {\cal{V}}_{M}(\tau)\,, \qquad  A_{adS}(\tau) \geq A_{M}(\tau)\,.
\ee
These inequalities are universal, valid for any values of $\tau$, and compare the volume of a causal diamond in a (anti-) de Sitter spacetime with that of in Minkowski spacetime.

\subsubsection{Geometric preliminaries}
The metric ansatz is the following
\begin{equation}
\label{nohads}
ds^2=-f(r)\e^{2\gamma(r)}dt^2 + f(r)^{-1}dr^2 +r^2d\Omega^2_{d-2}\, , \qquad f(r)=1-\frac{2m(r)}{r^{d-3}} - \widetilde{\Lambda}r^2 \, .
\end{equation}
where $\widetilde{\Lambda}=\frac{2\Lambda}{(d-1)(d-2)}$. The Einstein's equations with a cosmological constant $\Lambda$, $R_{\mu\nu}-\frac{1}{2}g_{\mu\nu}R +\Lambda g_{\mu\nu}=(d-2)\kappa_d T_{\mu\nu}$ produce the same equations for the functions $m(r)$ and $\gamma(r)$ as (\ref{md}) and (\ref{gmd}), and we also assume the same energy positivity condition as (\ref{T}), i.e. $T_{\hat{t}\hat{t}}\geq 0$, and  $T_{\hat{r}\hat{r}}\geq 0$, ensuring that the mass $m(r)$ is positive. At the origin $r=0$, we have that $f(r=0)=1$. 

The metric (\ref{nohads}) is asymptotically (anti-) de Sitter. If $\Lambda>0$ there exist a cosmological horizon $r_\Lambda$ but if $\Lambda<0$ there is not. The metric function $f(r)$ is non negative and only vanishes at $r=r_\Lambda$ if $\Lambda>0$,
\begin{align}
 f(r) > 0\,,\;\;  
\begin{cases}\displaystyle
r\in[0, r_\Lambda[ &\quad \text{if}\; \Lambda>0 \\ 
\displaystyle r\ge 0  & \quad \text{if}\; \Lambda<0
 \end{cases}
 \lb{fpos}
\end{align}
As follows from eq.~(\ref{gmd}) and (\ref{fpos}), $\gamma(r)$ is monotonic increasing function of radial coordinate $r$. The asymptotic boundary condition for $\gamma(r)$ is 
\be
\lim_{r\rightarrow r_\Lambda}\gamma(r)&=&0 \,, \qquad  \Lambda>0\\
\lim_{r\rightarrow \infty}\gamma(r)&=&0 \,, \qquad \Lambda<0
\lb{grads}
\ee
so that the metric component $g_{tt}=-f(r)e^{2\gamma(r)}$ approaches $-(1-\widetilde{\Lambda}r^2)$. Thus $\gamma(r)$ is negative and takes its minimal value at $r=0$, $\gamma(r=0)=\gamma_0<0$.

Let us see now if we have monotonicity of the red-shift. 
%
%
Looking at the derivative of $(tt)$-component of the metric with respect to $r$,
\be
-\frac{d g_{tt}(r)}{d r}=\frac{2e^{2\gamma(r)}}{r^{d-2}}\left((d-3)m(r)+\kappa_d\, r^{d-1} T_{\hat{r}\hat{r}} - \widetilde{\Lambda} r^{d-1} \right)\, ,
\lb{dgads}
\ee
we find that $-g_{tt}(r)$ is monotonically increasing function of $r$ in the case of a negative cosmological constant. On the other hand, if $\Lambda>0$ the term in parenthesis in the right hand side of (\ref{dgads}) is not sign definite and we cannot conclude on the monotonicity of $-g_{tt}(r)$. However, one can at least say that $-g_{tt}(r)$ is not a monotonically increasing function of $r$, as it would be in contradiction with the fact that $g_{tt}(0)=e^{2\gamma_0}>0$ and $g_{tt}(r_\Lambda)=0$. Therefore we will have to treat the cases $\Lambda>0$ and $\Lambda<0$ separately. 

Before doing so we give the general features of the causal diamond:

\noindent The volume of this diamond is
\be
V(\Lambda, \tau)=2 \Omega_{d-2}\int_0^{T/2}dt \int_0^{T/2-t} dy \, r^{d-2}(y) f(y)e^{2\gamma(y)} ,
\lb{V2}
\ee
and we will make use of its first derivative with respect to $\tau$ ($T=\tau e^{-\gamma_0}$)
\be
\frac{d V(\Lambda, \tau)}{d\tau}= \Omega_{d-2} \int_0^{r(T/2)}dr' r'^{d-2} e^{\gamma(r')-\gamma_0}\, .
\lb{dV1}
\ee
The area and the spatial volume are
\be
A(\Lambda, \tau)=\Omega_{d-2} r^{d-2}(T/2)\, , \qquad {\cal V}(\Lambda, \tau)=\Omega_{d-2}\int_0^{T/2}dy\, r^{d-2}(y)\sqrt{f(y)}e^{\gamma(y)}\, .
\lb{AV2}
\ee
The value of radial coordinate $r(T/2)$ is found from eq.~(\ref{T}) or, equivalently, in terms of invariant time interval from (\ref{tau1}). Since the mass term is positive, $m(r)>0$, we have that $e^{\gamma(r)} f(r)\le 1-\widetilde{\Lambda}r^2$, hence
\be
T/2=e^{-\gamma_0}\, \tau/2 \ge \int_0^{r(T/2)}\frac{dr'}{1-\widetilde{\Lambda}r'^2} \, .
\lb{rT2}
\ee
This inequality is universal valid for any value of $T$ and for any sign of $\Lambda$. 

\subsubsection{Comparison theorems for a positive cosmological constant}
With a positive cosmological constant $\widetilde{\Lambda} \equiv 1/\ell^2$, comparisons will be made with respect to de Sitter spacetime. From (\ref{rT2}) we find the inequality,
\be
r(T/2) \le \ell\tanh\frac{T}{2\ell}\,.
\lb{rTds}
\ee
Equality in (\ref{rTds}) correspond to de Sitter spacetime.

\paragraph{Inequality for volume of causal diamond}\mbox{}

From (\ref{dV1}) and using $e^{\gamma_0} \le e^{\gamma(r)}\le 1$ together with (\ref{rTds}) we have
\be
\frac{d}{d\tau}V(\Lambda>0, \tau) \leq \Omega_{d-2} e^{-\gamma_0}\int_0^{r(T/2)}dr' r'^{d-2} = \frac{\Omega_{d-2}}{d-1} e^{-\gamma_0}r^{d-1}(T/2) \leq  \frac{\Omega_{d-2}\ell^{d-1}}{d-1} e^{-\gamma_0}\tanh^{d-1}\frac{T}{2\ell}\,, \;\;
\lb{dVVds1}
\ee
where $\frac{\Omega_{d-2}\ell^{d-1}}{d-1}\tanh^{d-1}\frac{\tau}{2\ell}=dV_{dS}(\tau)/d\tau$. Integrating (\ref{dVVds1}) once with respect to $\tau$, we get the desired inequality,
\be
 V(\Lambda>0, \tau)\leq V_{dS}(\tau e^{-\gamma_0})\, ,
\lb{ineqVds}
\ee
where $V_{dS}(\tau)$ is the volume of the causal diamond in de Sitter spacetime.

\paragraph{Inequalities for area and spatial volume}\mbox{}

Concerning the area of causal diamond, we are also able to find an inequality involving the area of causal diamond in de Sitter spacetime, $A_{dS}(\tau)$. As follow straightforwardly from (\ref{rTds}), we have
\be
 A(\Lambda>0, \tau) \le A_{dS}(\tau e^{-\gamma_0})\,.
 \lb{ineqAds}
\ee

Conversely, we cannot find an inequality involving the spatial volume of de Sitter spacetime -- inequalities $f(r)\le 1-r^2/\ell^2$ and $e^{-\gamma(r)}\ge1$ and (\ref{rTds}) are not sufficient -- and we are only able to find 
\be
{\cal{V}}(\Lambda>0, \tau)\leq  e^{-(d-1)\gamma_0}\,\mathcal{V}_M(\tau) \, ,
\lb{ineqVspatds}
\ee
which is the upper bound in (\ref{spatV}).

\subsubsection{Comparison theorems for a negative cosmological constant}
With a negative cosmological constant $\widetilde{\Lambda} \equiv -1/\ell^2$, comparisons will be made with respect to anti-de Sitter spacetime. From (\ref{rT2})  we find the inequality
\be
 r(T/2) \le \ell\tan\frac{T}{2\ell}\,.
\lb{rTads}
\ee
Equality in (\ref{rTads}) correspond to anti-de Sitter spacetime.

\paragraph{Inequality for volume of causal diamond}\mbox{}

Using the same method as for the asymptotically de Sitter case (merely replacing hyperbolic tangent by the circular one through the change $\ell \rightarrow i\ell$), we find the following upper bound
\be
V(\Lambda<0, \tau) \le V_{adS}(\tau e^{-\gamma_0})\,,
\lb{ineqVads}
\ee
where $V_{adS}(\tau)$ is the volume of causal diamond of duration $\tau$ in anti-de Sitter spacetime.

\paragraph{Inequalities for area and spatial volume}\mbox{}

The inequality for the area reads
\be
 A(\Lambda<0, \tau) \le A_{adS}(\tau e^{-\gamma_0})\,,
\lb{ineqAads}
\ee
where we used (\ref{rTads}) and $A_{adS}(\tau)$ is the area of causal diamond in anti-de Sitter spacetime.

For the spatial volume we use $e^{\gamma(r)}\le 1$ and $f(r)\le 1+r^2/\ell^2$ and find
\be
{\cal{V}}(\Lambda<0, \tau)=\Omega_{d-2}\int_0^{T/2}dy\, r^{d-2}(y)\sqrt{f(y)}e^{\gamma(y)} \le \Omega_{d-2}\int_0^{T/2}dy\, r^{d-2}(y)\sqrt{1+r^2(y)/\ell^2}\,.
\ee
Now using (\ref{rTads}) we obtain the inequality
\be
{\cal{V}}(\Lambda<0, \tau) \le {\cal{V}}_{adS}(\tau e^{-\gamma_0}) \,,
\ee
where ${\cal{V}}_{adS}(\tau)$ is the spatial volume of the causal diamond in anti-de Sitter spacetime. 

We stress that the upper bounds we found for the volume, spatial volume, and area of diamond in our asymptotically (anti-) de Sitter spacetimes are universally  valid for any value of $\tau$.

\section{Towards generalization to arbitrary static metric}
\setcounter{equation}0
In the previous discussion the causal diamond was spherically symmetric. This is due to the fact that we considered
a spherically symmetric distribution of matter so that the spacetime in question was respecting this symmetry.
It is however not the most general case. In this section we make steps towards a generalization to include arbitrary static metric.
The idea is to extend all steps we have done in the spherical case to this more general spacetime.

We start with considering  a  general class of  static metrics,
\be
ds^2=-u^2(x)dt^2+h_{ij}(x)dx^idx^j\, .
\lb{g1}
\ee
On the hypersurface of constant $t$ one can always choose the appropriate normal system of coordinates in which the components of metric $h_{ij}$ take the form
\be
ds^2_{(d-1)}=h_{ij}(x)dx^idx^j=d\rho^2+\gamma_{ab}(\rho, \theta)d\theta^a d\theta^b\, ,
\lb{g2}
\ee
where $\rho$ is radial coordinate and $\theta^a\, , \ a=1,..,d-2$ are angular coordinates. This choice of the spatial coordinates will be useful below when we shall analyse some Bishop type inequalities.  

The other useful choice of spacetime coordinates is to introduce the tortoise coordinate $y$ or optical  radial  distance  in which the light cone structure is the easiest to analyse,
\be
ds^2=u^2(y,\theta)(-dt^2+dy^2)+\gamma_{ab}(y,\theta)d\theta^a d\theta^b\, .
\lb{g2-1}
\ee
First, we analyse whether the function $u(y,\theta)$ is in any sense monotonic.

\subsection{Monotonicity and the red-shift}
The Einstein equations in the metric (\ref{g1}) take the following form  (see e.g. \cite{Shiromizu}) 
\be
&&R_{00}=u \nabla ^2 u = S_{00}\, ,  \nonumber \\
&&R_{ij}= ^{(d-1)} R_{ij} - \frac{1}{u} \nabla _i \nabla _i u = S_{ij} \, ,
\lb{g2-2}
\ee
where $\nabla_i$ is the Levi-Civita covariant derivative of the metric
$h_{ij}$,  we have put
 $8 \pi G =1$. We introduced tensor
\be
S_{\mu \nu }&=&  T_{\mu \nu } - \frac{1}{d-2} g_{\mu \nu } 
T^\sigma _\sigma  \, .
\lb{g2-3}
\ee
So that one has that
\be
\nabla^2 u = u^{-1} \frac{1}{d-2} \Bigl
( (d-3) T_{00} +  T^k_k   \Bigr  )               \label{Poisson} \, .
\ee
Equation (\ref{Poisson}) may be considered as the general-relativistic
analogue of Poisson's equation in Newtonian gravity. We note that function $u(y,\theta)$ is subject to certain conditions.
First of all, since we consider the case without  horizons this function is everywhere positive,
\be
u(y,\theta)>0\  .
\lb{g2-4}
\ee
On the other hand, asymptotically, our metric (\ref{g1}) is supposed to approach the metric of Minkowski spacetime.  Therefore, we impose the boundary condition
at infinity
\be
u(y,\theta) \rightarrow 1 \, , \qquad y\rightarrow \infty\, .
\lb{u1}
\ee
It  now follows  from the {\it Strong  Energy Condition}, $S_{00} >0$, 
that 
\be
\nabla ^2 u \ge 0 \,.  \label{nomax}
\ee
The function satisfying equation (\ref{nomax}) is called {\it subharmonic}. The point now is that any {\it subharmonic}   function   $u(x)$ can have no local maximum.
This means that in any domain $\Omega$ the maximum of this function is reached on the boundary $\partial\Omega$.
This function  must have at least one local minimum. Suppose this minimum is  at point $y=y_c$ so that
\be
u(y=y_c,\theta)=u_c>0\, .
\lb{u3}
\ee
We can always adjust the coordinates such that $y_c=0$. 
We notice that the regularity requires that $u_c$ be independent of the angular coordinates $\{\theta\}$.
Let us now consider a ball of radius $Y>0$ with center at $y_c=0$.  By the maximum principle, the maximal value of function $u(y,\theta)\, , \  0<y\leq Y$, is reached on the boundary of the ball, i.e. at $y=Y$. Increasing $Y'>Y$ we consider a ball of larger radius $Y'$ which includes the smaller ball as a part. The maximum is always achieved on the boundary of the larger ball, i.e. at $y=Y'$. This property establishes a certain monotonicity of the function $u(y,\theta)$.
It should be noted that the angular coordinates of the maximum point change when $Y$ is increasing. Their exact values are not under our control.
Using this monotonicity, the asymptotic condition (\ref{u1}) as well as (\ref{u3}) we arrive at the inequality valid for any values of $y$ and for any values of angular coordinates
$\{ \theta \}$,
\be
u_c\leq u(y,\theta)<1\, .
\lb{u4}
\ee
The location of local   minimum $y=y_c$   would be the location of a timelike
geodesic observer,  $ y=y_c ={\rm const}$. 
The redshift $z_c$,  associated with the observer and defined by $1+z_c = 1/u_c $, 
must be greater than zero.  Using the inequalities (\ref{u4}) we conclude that for any other observer the red-shift
will satisfy the inequalities
\be
 0 < z \leq z_c\, 
 \lb{u5}
 \ee
 so that $z_c$ is the maximal red-shift in the spacetime. Equation (\ref{u5}), thus, generalizes the relations (\ref{zzc}) to the case of non-spherically symmetric spacetimes.

\subsection{Causal diamond, volume  and other geometric quantities}
We consider a causal diamond with the center at $y=y_c=0$.  For the coordinates of the points $p$ and $q$ we have that $p=(-T/2,0)$ and $q=(T/2, 0)$.
In the coordinate system (\ref{g2-1}) the light cone $I^-(q)$ is defined by equation $t=T/2-y$ while the light cone $I^+(p)$ is defined by
$t=-T/2+y$. The invariant duration of the diamond is 
\be
\tau=u_c \, T=T/(1+z_c)\, .
\lb{tau}
\ee
Equation $y=Y$ defines a closed surface of area
\be
A(Y)=\int d^{d-2}\theta\sqrt{\gamma(Y,\theta)}u(Y,\theta)\, .
\lb{u6}
\ee
Integrating this area with respect to $y$ we obtain the spatial $(d-1)$-volume on the hypersurface of constant time $t$,
\be
{\cal{V}}_{d-1}(Y)=\int_0^Y dy\, A(y)\, .
\lb{u7}
\ee
The volume of the causal diamond is then  obtained by doing one more integration,
\be
V(T)=2\int_0^{T/2}\, dt\, {\cal{V}}_{d-1}(T/2-t)\, .
\lb{u8}
\ee
Using the inequality (\ref{u4}) we immediately arrive at the following inequalities for the area $A(y)$,
\be
\frac{1}{1+z_c}\bar{A}(y)=u_c\bar{A}(y)\leq A(y)\leq \bar{A}(y)\, ,
\lb{u9}
\ee
where
\be
\bar{A}(y)=\int d^{d-2}\theta \sqrt{\gamma(y,\theta)}
\lb{u10}
\ee
is the area in the metric (\ref{g2-1}) with $u=1$.

\subsection{Bishop's type inequalities}
Now we would like to derive the Bishop's type inequality for the area (\ref{u10}), defined in the metric (\ref{g2-1}) with $u=1$ or, effectively, in the spatial metric (\ref{g2}).
Then, provided we are successful, this inequality can be further integrated once or twice to obtain the inequalities for the spatial volume (\ref{u7}) or the volume of the causal diamond
(\ref{u8}).

The Ricci tensor of the spatial metric (\ref{g2}) reads
\be
&&^{(d-1)}R_{ab}=^{(d-2)}R_{ab}(\gamma)-\frac{1}{2}\partial_\rho^2\gamma_{ab}+\frac{1}{2}(\partial_\rho\gamma\gamma^{-1}\partial_\rho\gamma)_{ab}-\frac{1}{4}\partial_\rho\gamma_{ab}\Tr(\gamma^{-1}\partial_\rho\gamma)\, , \nonumber \\
&&^{(d-1)}R_{\rho\rho}=-\frac{1}{2}\partial_\rho\Tr(\gamma^{-1}\partial_\rho\gamma)-\frac{1}{4}\Tr(\gamma^{-1}\partial_\rho\gamma)^2\, 
\lb{g3}
\ee
and the Ricci scalar
\be
^{(d-1)}R=^{(d-2)}R(\gamma)-\partial_\rho\Tr(\gamma^{-1}\partial_\rho\gamma)-\frac{1}{2}(\tr(\gamma^{-1}\partial_\rho\gamma))^2\, ,
\lb{g4}
\ee
where $^{(d-2)}R_{ab}(\gamma)$ and $^{(d-2)}R(\gamma)$ is respectively the Ricci tensor and the Ricci scalar of $(d-2)$-metric $\gamma_{ab}$.
The surface $\Sigma$ of radius $\rho$ has the area
\be\qquad
\bar{A}(\rho)=\int_\Sigma d^{d-2}\theta \sqrt{\gamma(\rho,\theta)}\, ,  \qquad \gamma(\rho,\theta)=\det \gamma_{ab}(\rho, \theta)\, .
\lb{g5}
\ee
Being inspired by our analysis in Section \ref{Bishop} we define 
\be
\theta=\bar{A}^{-1}\frac{d\bar{A}}{d\rho}=\frac{1}{2\bar{A}}\int_\Sigma \sqrt{\gamma}\Tr(\gamma^{-1}\partial_\rho\gamma)\, .
\lb{g6}
\ee
The derivative of this quantity with respect to $\rho$ can be evaluated as follows
\be
\frac{d\theta}{d\rho}=-\theta^2+\frac{1}{2\bar{A}}\int_\Sigma(\partial_\rho\Tr(\gamma^{-1}\partial_\rho\gamma)+\frac{1}{2}(\tr(\gamma^{-1}\partial_\rho\gamma))^2)
\ee
We can use either equation (\ref{g3}) or (\ref{g4}) to express the second derivative of $\gamma$ in terms of the curvature. We then find the following
form for this equation,
\be
\frac{d\theta}{d\rho}=-\theta^2+\frac{1}{\bar{A}}\int_\Sigma\left( -R_{\rho\rho}+\frac{1}{4}(\Tr(\gamma^{-1}\partial_\rho\gamma))^2-\frac{1}{4}\Tr(\gamma^{-1}\partial_\rho\gamma)^2                       \right)\, .
\lb{g7}
\ee
Matrix $\Gamma=(\gamma^{-1}\partial_\rho\gamma)$ can be decomposed on the trace and the traceless parts,
\be
(\gamma^{-1}\partial_\rho\gamma)^a_b=\frac{\delta^a_b}{d-2}\Tr(\gamma^{-1}\partial_\rho\gamma)+\tilde{\Gamma}^a_b\, , \qquad \Tr\tilde{\Gamma}=0\, .
\lb{g8}
\ee
Using this expansion we obtain for the equation (\ref{g7})
\be
\frac{d\theta}{d\rho}=-\theta^2+\frac{1}{\bar{A}}\int_\Sigma\left( -R_{\rho\rho}-\frac{1}{4}\Tr\tilde{\Gamma}^2+\frac{1}{4}(\frac{d-3}{d-2})(\Tr(\gamma^{-1}\partial_\rho\gamma))^2\right)\, .
\lb{g9}
\ee
We have seen above that $-R_{\rho\rho}$ is positive, at least outside the matter source. On the other hand, $-\frac{1}{4}\Tr\tilde{\Gamma}^2$ contributes negatively to the right hand side of (\ref{g9}). Notice that in the spherically symmetric case the traceless part of $\Gamma$ vanishes and we can formulate a lower bound for $\frac{d\theta}{d\rho}$, as was discussed in section \ref{Bishop}.
In order to proceed same way  in a non-spherically symmetric case we need a new, combined, positivity condition. Thus, we assume that
\be
-R_{\rho\rho}-\frac{1}{4}\Tr\tilde{\Gamma}^2\geq 0
\lb{g10}
\ee
everywhere on the hypersurface of constant time $t$.  Combining this condition with the inequality
\be
\int_\Sigma (\Tr(\gamma^{-1}\partial_\rho\gamma))^2\geq \frac{1}{\bar{A}}\left(\int_\Sigma \Tr(\gamma^{-1}\partial_\rho\gamma)\right)^2
\lb{g11}
\ee
valid for any function $\Tr(\gamma^{-1}\partial_\rho\gamma)$, we arrive at the inequality
\be
\frac{d\theta}{d\rho}\geq -\frac{\theta^2}{d-2}\, .
\lb{g12}
\ee
Representing  $\bar{A}=\Omega_{d-2}w^{d-2}$  and hence  $\theta=(d-2)w^{-1}\partial_\rho w$ we
find 
\be
\partial_\rho^2 w\geq 0\, .
\lb{g13}
\ee
Integrating this inequality from $\rho$ to $\infty$ and using the asymptotic condition  that $w\rightarrow \rho$ if $\rho\rightarrow \infty$ we obtain
\be
\partial_\rho w\leq 1 \Rightarrow w\leq 1\, .
\lb{g14}
\ee
The latter inequality suggests that
\be
\bar{A}(\rho)\leq \Omega_{d-2}\rho^{d-2}=A_M(\rho)\, .
\lb{g15}
\ee
This inequality can be integrated with respect to $\rho$ from $0$ to $Y$ and we obtain the Bishop's inequality for the spatial volume 
\be
\bar{{\cal{V}}}(Y)\leq {\cal{V}}_M(Y)
\lb{g16}
\ee
computed for the metric (\ref{g2-1}) with $u=1$, ${\cal{V}}_M(Y)$ is the volume in Minkowski spacetime.

\subsection{Inequality for volume of causal diamond}

The relations (\ref{g15})  should be combined with (\ref{u9}) to get
\be
A(Y)\leq A_{M}(Y)\, ,
\lb{g17}
\ee
where $A(Y)$ is the area computed in the metric (\ref{g2-1}) with non-constant $u$.   The integration of this inequality with respect to $Y$ then gives the inequality
for the spatial volume (\ref{g7}) computed in the complete metric (\ref{g2-1}),
\be
{\cal{V}}(Y)\leq {\cal{V}}_M(Y)\, ,
\lb{g18}
\ee
where ${\cal{V}}_M(Y)$ is the volume computed in Minkowski spacetime. One more integration gives us the comparison theorem for the volume of causal diamond,
\be
V(T)\leq V_M(T)\, 
\lb{g19}
\ee
or, equivalently, expressing this relation in terms of the invariant duration $\tau$ (\ref{tau}), we arrive at relation (\ref{V<M}),
\be
V(\tau)\leq (1+z_c)^dV_M(\tau)\, .
\lb{g20}
\ee
This completes our generalization of the comparison theorems to non-spherical causal diamonds and generic static spacetime.

\section{Diamonds in spacetime with horizon} 
\begin{figure}[ht]
\begin{center}
\includegraphics[height=4.5cm]{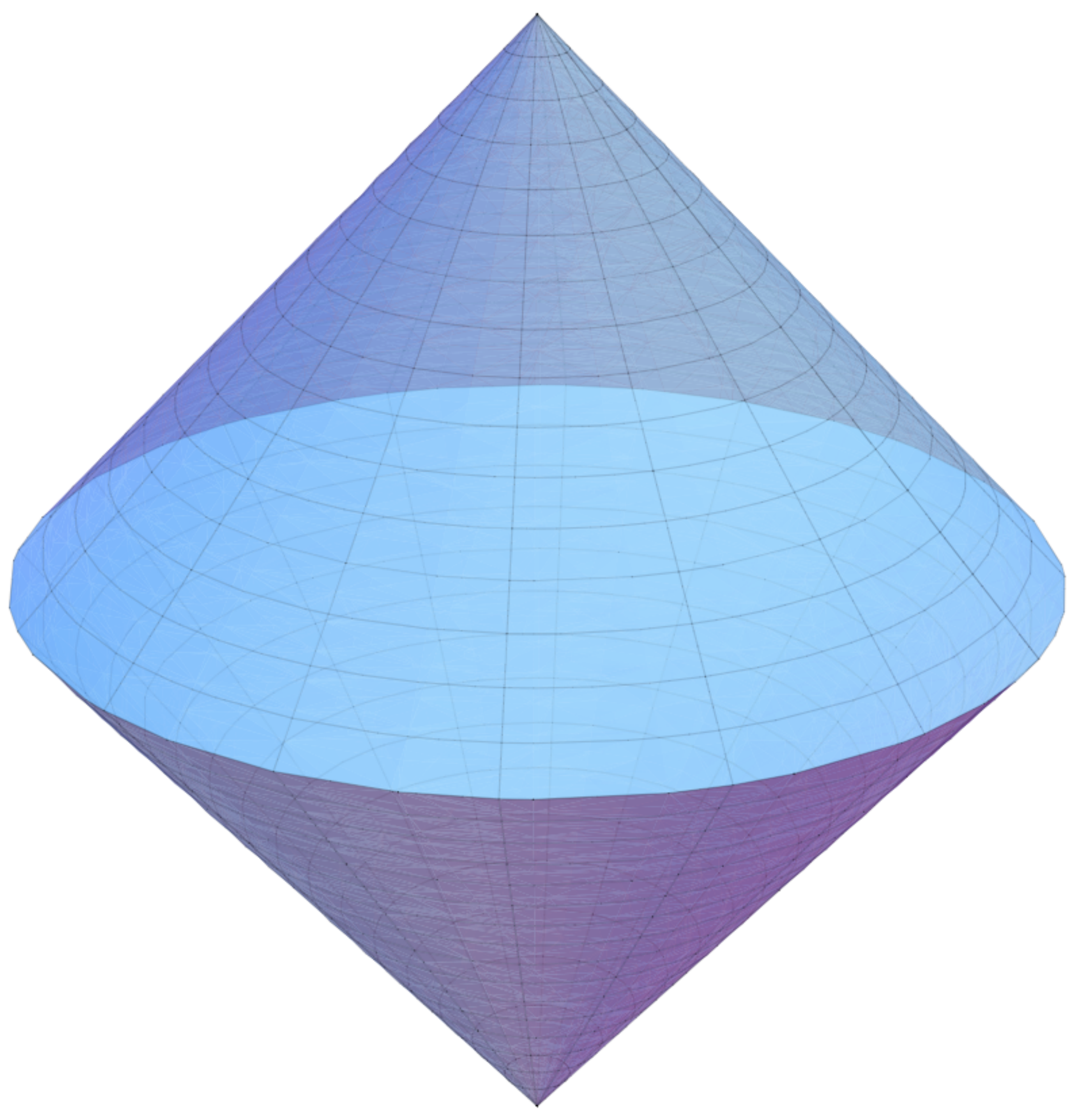}\hspace{0.8cm}
\includegraphics[height=4.5cm]{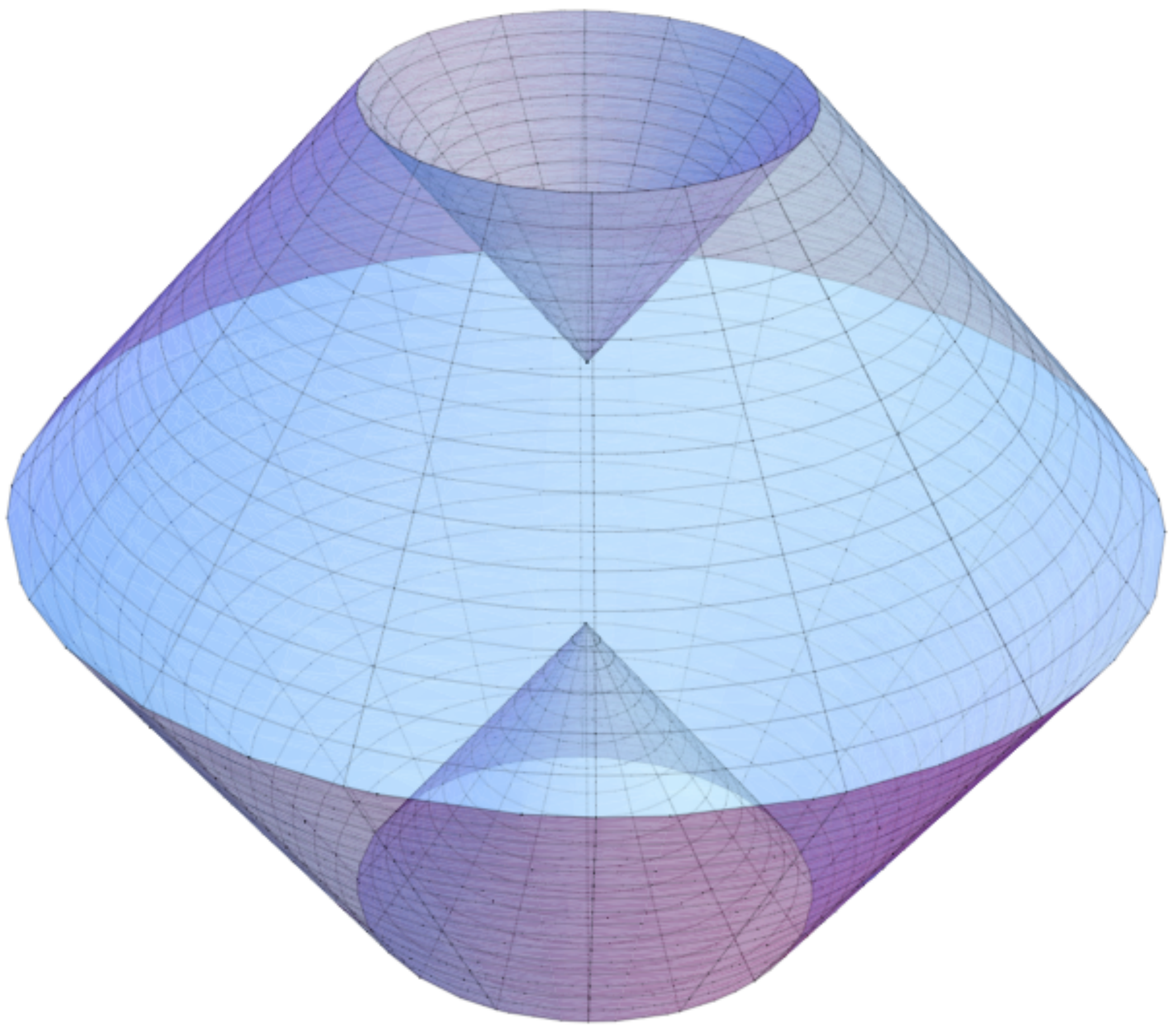}\hspace{0.8cm}
\includegraphics[height=4.3cm]{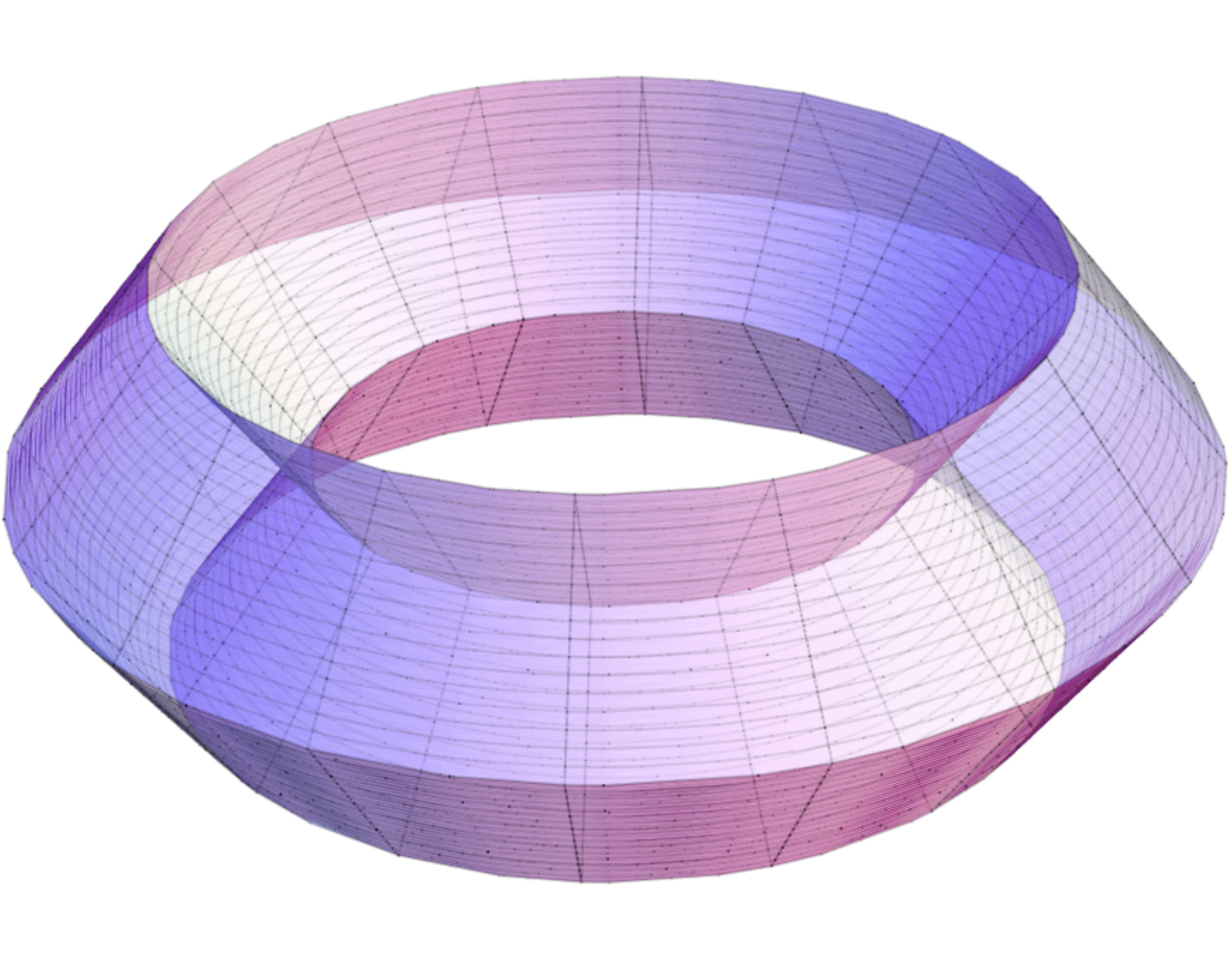}
\end{center}
\caption{Diamonds in $2+1$ dimensions:  unshifted center (left) and shifted center (center) in Minkowski spacetime, shifted center in spacetime with a horizon (right).}
\label{figdiam}
\end{figure}

\setcounter{equation}0 
In this section we are interested in a situation when a black hole
horizon is present in the spacetime. For simplicity and for the sake
of concreteness we shall consider the case of the $d$-dimensional
Schwarzschild metric. The causal diamonds in this spacetime become
rather complicated objects since they are not spherically symmetric,
the light cones emitted from points $p$ and $q$ are essentially
dependent on the angular coordinates. Since we want to consider a
simpler situation when we still can use the spherical symmetry we
shall generalize the notion of the causal diamond.   Let us consider
a diamond  with a shifted center (see
\hyperref[figdiam]{Figure~\ref*{figdiam}}). In this case $p$ and $q$
are not points but $(d-2)$-spheres.

These  generalised type of diamonds may be understood in 
terms of the notion of {\it domain
of dependence}. For an achronal set $S$ , for example  a domain
in a surface of constant time in a static spacetime, the {\it future
domain of dependence $D^+(S)$} is the set of points $p$ such that every past in
extendible acausal  curve intersects $S$ once and only once. There is 
a temporally dual  notion of a  {\it past   domain of dependence  $D^-(S)$}.
One may also  consider the union $D(S)= D^+(S)\cup D^-(S)$. A causal diamond
in Minkowski spacetime may be thought of $D(S)$ where $S$ is a ball in a 
spacelike  hypersurface which is orthogonal to the timelike geodesic joining
$p$ and $q$ and this is  also true of the causal diamonds we
have been considering in spherically symmetric spacetimes. 
The more general diamonds we are considering in this section may also
be considered as domains of dependence.  The achronal set in this case
is a solid  annulus of the form $I \times S^{d-2} $, where
$I$ is the interval in the radial direction given by 
$ y_0- \half \tau  \le y_0 + \half \tau$, where
the diamond has its center  at the  tortoise coordinate $y=y_0$.
If a black hole is present, then $ y_0- \half \tau $ will always 
lie out side the horizon since this is situated at $y=-\infty$. 
If no black hole is present than we only obtain an annulus
if  $ y_0- \half \tau $ is positive (assuming that the centre of spherical
symmetry is  at $y=0$). If   $ y_0- \half \tau <0$ , the domain is a ball
as illustrated  on the left hand side of figure 1.

For a fixed geometry, the volume of this diamond is completely determined by only two variables: the invariant duration $\tau$ and the position of its center $r_0$. In $d$-dimensional Schwarzschild spacetime, with the following metric
\be
ds_{Sch^d}^2=-f_d(r)dt^2 + f_d(r)^{-1}dr^2 +r^2d\Omega^2_{d-2}\, , \qquad \ f_d(r)=1-\Big(\frac{r_s}{r}\Big)^{d-3}\,  , 
\ee
the volume of such a diamond is
\be
\label{vsch}
V_{Sch^d}(T,r_0) = 2 \Omega_{d-2}\int_{0}^{T/2}d t \int_{-T/2+t}^{T/2-t}d y\, r^{d-2}(y)f_d(y) \, ,
\ee
where the tortoise coordinate $y = \int dr\, f_d^{-1}(r)$ and 
\be
\label{ysch}
T/2 = \int_{r_0}^{r(T/2)} \frac{dr'}{f_d(r')} = y(r(T/2))-\alpha_d(r_0) \,.
\ee
where $\alpha_d(r_0)\equiv y(r_0)$ and $r_0>r_s$. The volume can be displayed in an integrated form. First, we integrate over $y$ which gives
\begin{equation}
V_{Sch^d}(T,r_0) = \frac{2 \Omega_{d-2}}{d-1} \int^{T/2}_0 d t\; \left(r^{d-1}(T/2-t)-r^{d-1}(t-T/2)\right)\,.
\end{equation}
Performing the integration over $t$ we defined 
\begin{equation}
\label{Fd}
\int d t\;r^{d-1}(\alpha \pm t) = \pm \int d r\, \frac{r^{d-1}}{1-(r_s/r)^{d-3}} = \pm F_d(r(\alpha \pm t))
\end{equation}
and find for the volume
\begin{equation}
\label{vschint}
V_{Sch^d}(T,r_0) = \frac{2\Omega_{d-2}}{d-1} \Big(F_d(r(T/2)) +F_d(r(-T/2))-2F_d(r_0) \Big)\,.
\end{equation}

In above formulas $T$ is the duration of the diamond measured in the clock of a distant observer at the spatial infinity. this duration is different from the actual, geometric, duration of the diamond measured by an observer which travels from $p$ to $q$.  
We notice, however, that  since $r={\rm const} \ne 0$ 
is not geodesic,   this world line is not the central  geodesic
of the generalized causal diamond and the duration
$\tau$ of these causal diamonds is \emph{longer than}  
$T \sqrt{f_d(r_0)}$. In fact we will see that one can derive an inequality for the function $T(\tau)$, valid for all dimensions, involving the red-shift at the center $r_0$ of the shifted diamond. In what follows we shall analyse the exact relation between $T$ and $\tau$.

Before going any further we want to compute (\ref{ysch}) and find the relation between $T$ and $\tau$ the proper time along the radial geodesic $(p, q)$.

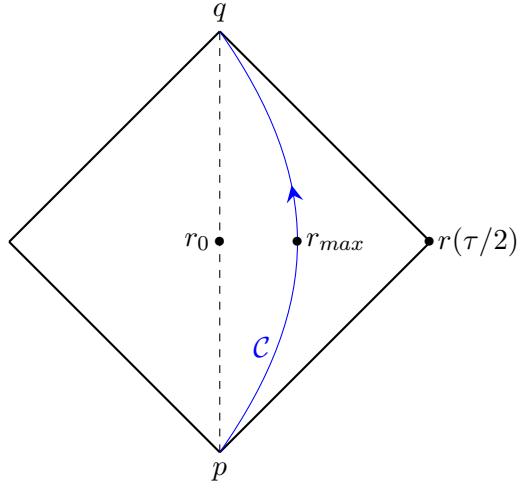
\begin{figure}[h]
\begin{center}
\begin{tikzpicture}[scale=1.4]

\draw[thick] (-2,0) --+ (2,2);
\draw[thick] (0,2) --+ (2,-2);
\draw[thick] (-2,0) --+ (2,-2);
\draw[thick] (0,-2) --+ (2,2);
\draw[dashed] (0,-2) --+ (0,4);

\draw[blue,decoration={markings, mark=at position 0.625 with {\arrow[scale=2]{stealth}}},
        postaction={decorate}] (0,-2) to[out=55, in=-55, looseness=1.1] (0,2);
\node [right,blue] at (0.2,-1) {$\mathcal{C}$};

\node [left] at (0.03,0) {$r_0$};
\node[scale=0.8] at (0,0) {$\bullet$};
\node[scale=0.8] at (0.735,0) {$\bullet$};
\node[scale=0.8] at (1.99,0) {$\bullet$};
\node [right] at (1.95,0) {$r(\tau/2)$};
\node [right] at (0.7,0) {$r_{max}$};
\node [above] at (0,2) {$q$};
\node [below] at (0,-2) {$p$};

\end{tikzpicture}
\vspace{-10pt}
\end{center}
\caption{Schematic of causal diamond in Schwarzschild spacetime. An observer travels from $p$ to $q$ along a radial geodesic $\mathcal{C}$ with $r_{max}$ being the apogee of the path and $r_0$ the center of the diamond. The radius of the diamond is $r(\tau/2)$. The geodesic $\mathcal{C}$ remains inside the diamond.}
\end{figure}

\subsection{Tortoise coordinates and radial timelike geodesic in $d$-dimensions}
\subsubsection{A useful integral}
In our calculations, the following integral will be needed,
\be
R_d(x) = \int \frac{dx}{x^{d-3}-1}
\ee 
This integral can be computed explicitly and one finds \cite{Gradshteyn}
\be
\label{Re}
R_d(x) = \frac{1}{d-3}\ln(x-1)- \frac{1}{d-3}\sum_{k=0}^{\frac{d-6}{2}}\left(P_k\,\cos\frac{(2k+1)\pi}{d-3} + Q_k\,\sin\frac{(2k+1)\pi}{d-3}\right)\, ,
\ee
for $d$ even, where $\displaystyle P_k(x) = \ln\Big(x^2+ 2x \cos\frac{(2k+1)\pi}{d-3} +1\Big),\;\; Q_k(x) = 2\arctan\left(\frac{x+\cos\frac{(2k+1)\pi}{d-3}}{\sin\frac{(2k+1)\pi}{d-3}}\right) -\pi$, and
\be
\label{Ro}
R_d(x) = \frac{1}{d-3}\ln\frac{x-1}{x+1} + \frac{1}{d-3}\sum_{k=1}^{\frac{d-5}{2}}\left(P'_k\,\cos\frac{2k\pi}{d-3} - Q'_k\,\sin\frac{2k\pi}{d-3}\right)\, ,
\ee
for $d$ odd, where $\displaystyle P'_k(x) = \ln\Big(x^2- 2x \cos\frac{2k\pi}{d-3} +1\Big),\;\; Q'_k(x) = 2\arctan\left(\frac{x-\cos\frac{2k\pi}{d-3}}{\sin\frac{2k\pi}{d-3}}\right)-\pi\,$.

\subsubsection{Tortoise coordinates}
The indefinite integral $y(r)$ in eq. (\ref{ysch}) can be rewritten as
\be
\label{ysch2}
y(r) = \int \frac{dr}{1-(r_s/r)^{d-3}} =  r + r_s\, R_d(r/r_s) \,.
\ee
For large $r$, we have to distinguish two cases. In $d=4$ dimensions, (\ref{ysch2}) reduces to
\be
\label{yinf4}
y(r\rightarrow +\infty) \simeq r + \,r_s\ln r/r_s - \mathcal{O}(r^{-1})  \, ,
\ee
and for $d>4$ we have
\be
\label{yinfd}
y(r\rightarrow +\infty) \simeq r - \mathcal{O}(r^{-(d-4)})  \, .
\ee
Near the horizon $r_s$ we find that
\be
y(r\rightarrow r_s) \simeq \frac{r_s}{d-3}\ln(r/r_s-1) + \mathcal{O}(1)\, . 
\lb{yrs}
\ee
valid for all dimensions.

\subsubsection{Proper time of radial timelike geodesics}

In $d$-dimensional Schwarzschild spacetime, a timelike radial geodesic obeys the following equations,
\begin{eqnarray}
&& \frac{dt}{d\tau} = \frac{E}{f_d(r)}\, , \label{teq} \\
&& \Big(\frac{dr}{d\tau}\Big)^2 = E^2 - f_d(r) \, , \label{req}
\end{eqnarray}
where $\tau$ is the proper time and $E^2=f_d(r_{max})$ is a constant of motion with $r_{max}$ being the apogee of the path for a bounded trajectory from $r_0$ to $r_{max}$ and then back to $r_0$. Our diamond is determined by two events $p$ and $q$ (of coordinates $p=(-T/2,r_0)$ and $q=(T/2, r_0)$) which are joined by a radial timelike geodesic. Since we are interested in the volume of large diamonds, we will have take $r_{max} \rightarrow + \infty$. Our goal here is to link the duration $T$ to the invariant duration $\tau$. 

Before going into the details, one can find some nice inequalities on $T(\tau)$. Throughout the round trip we have 
\be
f_d(r_0) \leq f_d(r) \leq E^2 < 1\,.
\lb{Eineq}
\ee
From the geodesic equation (\ref{teq}) it follows straightforwardly, using (\ref{Eineq}),
\be
\tau < T \leq (1+z_0)^2\,\tau \,.
\lb{Ttauineq}
\ee
where $1/(1+z_0)^2=f_d(r_0)$ such that $0< z_0<+\infty$. 
We notice that this inequality is universal, it is valid for any dimension $d\geq 4$, and involves the red-shift at the center $r_0$ of the shifted diamond. However, we will need to know in more detail the asymptotic behavior of function $T(\tau)$ in various dimensions.

Let us now get down to the details and find what we are looking for, that is the function $T(\tau)$. 
We start by considering the round trip from $r_0$ to $r_{max}$ and then back to $r_0$, with $r_{max}$ finite and $r_0>r_s$ fixed. Since the trajectory is symmetric, only half of this trip is needed to be considered, say from $r_0$ to $r_{max}$. 
Thus $r_{max}$ and $r_0$ fix the duration $\tau$, or equivalently, $\tau$ and $r_0$ fix the value of $r_{max}$. Therefore, $E(r_{max})$ is completely determined by $\tau$ ($r_0$ is fixed). We use eq.~(\ref{req}) to find $r_{max}$ as a function of $\tau$,
\be
\int_0^{\tau/2}d\tau' = \tau/2 = \int_{r_0}^{r_{max}}\frac{dr}{\sqrt{E^2-f_d(r)}}\,,
\ee
which, in the approximation of large $r_{max}$ (i.e. large $\tau$), gives 
\be
r_{max}(\tau) \simeq r_s \Big(\frac{\tau}{2b_d r_s} \Big)^{2/(d-1)} \,,
\ee
where $\displaystyle b_d = \sqrt{\pi}\, \frac{\Gamma\left((d-1)/(2(d-3))\right)}{\Gamma\left(1/(d-3)\right)}$. Then we can express $E$ in terms of $\tau$, 
\be
E^2(\tau) \simeq 1 - \Big(\frac{2b_d r_s}{\tau} \Big)^{\frac{2(d-3)}{d-1}}\,.
\ee
Now eqs.~(\ref{teq}) and (\ref{req}) should be combined,
\be
T/2 = \int_{r_0}^{r_{max}(\tau)}dr \frac{E(\tau)}{f_d(r)\sqrt{E^2(\tau)-f_d(r)}}\,,
\lb{Trmax2}
\ee
giving us the duration $T$ as a function of $\tau$ (and $r_0$). In the large $\tau$ approximation one obtains
\be
\label{Ttau}
T/2  \,\simeq\, \tau/2 + \frac{a_d\,r_s}{5-d} \Big(\frac{\tau}{2r_s} \Big)^{\frac{5-d}{d-1}}  + \beta_d(r_0) \, ,
\ee
where $a_d =\frac{1}{2} (3d-7) b_d^{2(d-3)/(d-1)}$ and $\beta_d(r_0) > 0$ is a constant depending on $r_0$. 
We notice that the behavior of function $T(\tau)$ is different in the cases $d=4$, $d=5$ and $d>5$. In dimension $d=4$ the correction term in (\ref{Ttau}) is a positive 
power of $\tau$ with a positive pre-factor. In dimension $d>5$ this pre-factor becomes negative but also the power of $\tau$ becomes negative. Therefore, asymptotically for large $\tau$ the first correction term would be constant, $\beta_d(r_0)>0$.  In dimension $d=5$ the power law in the correction term becomes a logarithm, 
\be
T/2\,\simeq\, \tau/2+\frac{r_s}{2}\ln\tau/r_s+\ ...\ .
\lb{d=5}
\ee

\subsection{Volume in terms of $T$}
\lb{secr0}
We first want to find an (asymptotic) inequality on the volume in terms of $T$. To do so we have to evaluate the volume given eq.~(\ref{vschint}) for large $T$; that is to say find the asymptotic behavior of the function $F_d(r(T/2))$ defined in (\ref{Fd}). Simple calculations show that 
\be
F_d(r) \simeq \frac{r^d}{d} + r_s^{d-3}\frac{r^3}{3} + \cdots \,.
\lb{Fdap}
\ee
Next we need the asymptotics of the function $r(\pm T/2)$. In fact we can discard $r(-T/2)$ because it is $\mathcal{O}(1)$.  We will treat the cases $d=4$ and $d>4$ separately.

\subsubsection{The case $d=4$}
Using (\ref{ysch}) and (\ref{yinf4}) we obtain the asymptotic of the function $r(T/2)$,
\be
 \label{rp}
r(T/2) &\simeq& \frac{T}{2} -r_s\ln\frac{T}{2r_s} + \mathcal{O}(1)  \, .
\ee
Because of the logarithm term, in $d = 4$  we have that
\be
r(T/2) < \frac{T}{2} \, ,
\ee
hence the volume of a large diamond,
\be
V_{Sch^4}(T, r_0) \simeq \frac{\pi}{24}T^4 -  \frac{\pi}{3}r_s T^3 \ln\frac{T}{2r_s} + \mathcal{O}(T^3)  <  \frac{\pi}{24}T^4 = V_{M^4}(T)\, ,
\ee
is always smaller than that of one in Minkowski spacetime, no matter
whether we  consider  a diamond with a shifted center or not. This statement is independent of the position $r_0$ of the center of the diamond.

\subsubsection{The case $d>4$}

In dimensions $d > 4$ the situation is different. There is no logarithmic term in $r(T/2)$, 
\be
r(T/2) \simeq \frac{T}{2} + \alpha_d(r_0) + \mathcal{O}(r^{-(d-4)}) \, ,
\ee
and the volume takes the form
\be
V_{Sch^d}(T, r_0) \simeq \underbrace{\frac{\Omega_{d-2}}{2^{d-1} d (d-1)}T^d}_{\displaystyle =V_{M^d}(T)} + \;\alpha_d(r_0)\frac{\Omega_{d-2}}{2^{d-2}(d-1)}T^{d-1}  + \mathcal{O}(T^{d-2}) \,.
\ee
Thus the volume of the diamond can be smaller or larger than the volume in Minkowski spacetime, depending on the sign of $\alpha_d(r_0)$.
Indeed, there exists some critical value $r_c$ such that for $r_0 < r_c$ we have $V_{Sch^d}(T, r_0)<V_{M^d}(T)$ (i.e. if $\alpha_d(r_0)<0$) and for $r_0 > r_c$ we have $V_{Sch^d}(T, r_0)>V_{M^d}(T)$ (i.e. if $\alpha_d(r_0)>0$). However, as follows from the table below, these critical values $r_c$ correspond to values of $r_0$ which are very close to the horizon $r_s$. Interestingly, the critical value $r_c$ in dimension $d > 4$ is a new important radius which characterizes the Schwarzschild spacetime.
\vspace{5pt}

\begin{table}[h]
\begin{center}
\begin{tabular}{|c|ccccccc|}
\hline
 Dimension  & $d = 5$ & $d = 6$ & $d = 7$ & $d = 8$ & $d = 9$ & $d = 10$ & $d = 11$ \\ 
\hline
 $\,r_c/r_s$ & $1.19968 $ & $1.03261$ & $1.00747$ & $1.00197$ & $1.00056$ & $1.00017$ & $1.00005$ \\
 \hline
\end{tabular}
\vspace{-10pt}
\end{center}
\caption{Critical values $r_c$ ($\alpha_d(r_c)=0$) of $r_0$}
\label{r0}
\end{table}

Summarizing, we see that the inequality
\begin{equation}
V_{Sch^d}(T, r_0)<V_{M^d}(T)\, ,
\end{equation}
is always valid  in dimension $d=4$. In dimension $d>4$ it is valid only if $r_0$ is very close to horizon.
Otherwise, for a ``typical'' observer, we have that
\begin{equation}
V_{Sch^d}(T, r_0)>V_{M^d}(T)\, , \qquad \text{for} \quad d>4
\end{equation}

\subsection{Volume in terms of $\tau$}
In the previous subsection we have obtained an inequality on the volume of a diamond in Schwarzschild spacetime in terms of $T$. Now we want to analyse it  in terms of the invariant duration $\tau$, the proper time associated with a radial geodesic joining spheres $p$ and $q$.
 We recall
\begin{eqnarray}
\lb{vsch4}
V_{Sch^4}(T, r_0) &\simeq& \frac{\pi}{24}T^4 -  \frac{\pi}{3}r_s T^3 \ln\frac{T}{2r_s} + \mathcal{O}(T^3) \, , \\
V_{Sch^d}(T, r_0) &\simeq& V_{M^d}(T) \left(1 + 2d \frac{\alpha_d(r_0)}{T}  + \mathcal{O}(T^{-2})\right) \,.
\lb{vschd}
\end{eqnarray}
where $\displaystyle V_{M^d}(T)$ is the volume of a causal diamond in $d$-dimensional Minkowski spacetime. We already know from (\ref{Ttauineq}) that $T > \tau$.  However,  this inequality is not sufficient to draw any firm conclusion. 
In dimension $d=4$ the reason is that the correction term in (\ref{vschd}) is negative. Therefore, we cannot use $T>\tau$ directly. Instead, one has to analyse carefully the subleading terms in
the asymptotic expansion for large $\tau$. In dimension $d>4$ the correction term in (\ref{vschd}) is not sign definite. The sign depends on the value of $r_0$, as we discussed this in 
section \ref{secr0}. 

We have to distinguish three cases : $d=4$ and $d=5$ and $d>5$.

\subsubsection{The case $d=4$}
In $d=4$ dimensions we have 
\be
T/2  \,\simeq\, \tau/2 + \frac{5 r_s}{2}\Big(\frac{\pi^2\tau}{8r_s}\Big)^{1/3} + \mathcal{O}(1) \, ,
\label{Ttau4}
\ee
as follow from eq.~(\ref{Ttau}). Injecting the equation above in the asymptotic formula of the volume (\ref{vsch4}) one gets 
\be
V_{Sch^4}(\tau, r_0) &\simeq& \frac{\pi}{24}\tau^4 + \frac{5\pi r_s}{6}\Big(\frac{\pi^2}{8r_s}\Big)^{1/3} \tau^{10/3} -  \frac{\pi}{3}r_s \tau^3 \ln\frac{\tau}{2r_s} \,>\, \frac{\pi}{24}\tau^4 = V_{M^4}(\tau) \,.
\lb{Vschshift4}
\ee
The logarithmic term is now subdominant since
\be
\tau^{10/3} > \tau^3\ln\tau\,, \qquad \tau \rightarrow +\infty\,,
\ee
and we conclude that the volume in terms of $\tau$ is always larger than that of in Minkowski spacetime,
\be
V_{Sch^4}(\tau, r_0) > V_{M^4}(\tau) \,.
\ee

\subsubsection{The case $d=5$}
In $d=5$ dimensions the relation between $T$ and $\tau$ reads 
\be
\label{Ttau5}
T/2  \,\simeq\, \tau/2 + \frac{r_s}{2}\ln \tau/r_s + \mathcal{O}(1) \,,
\ee
and we find for the volume
\be
V_{Sch^5}(\tau, r_0) &\simeq& \frac{\pi}{160}\tau^5 + \frac{\pi^2}{32}r_s\tau^4\ln\tau/r_s + \mathcal{O}(\tau^4) \,>\, \frac{\pi}{160}\tau^5 = V_{M^5}(\tau) \,.
\lb{Vschshift5}
\ee
As in the 4-dimensional case, we find the inequality
\be
V_{Sch^5}(\tau, r_0) > V_{M^5}(\tau) \,.
\ee

\subsubsection{The case $d>5$}
Finally, for $d>5$ dimensions, the next to leading order in the equation relating $T$ to $\tau$ is of order $\mathcal{O}(1)$, that is
\be
T/2  \,\simeq\, \tau/2  + \beta_d(r_0) - \mathcal{O}(\tau^{-\frac{d-5}{d-1}})  \,.
\lb{Ttaud}
\ee
Therefore, using (\ref{vschd}) together with (\ref{Ttaud}) we find that the volume as a function of $\tau$ is
\be
V_{Sch^d}(\tau, r_0) &\simeq& V_{M^d}(\tau) \left(1+  2d \frac{\sigma_d(r_0)}{\tau}  + \mathcal{O}(\tau^{-2})\right) \,,
\ee
where $\sigma_d(r_0) = \alpha_d(r_0) + \beta_d(r_0)$ is a constant. The sign of the constant $\sigma_d(r_0)$ is crucial as it will dictate whether the inequality we find is a lower bound or an upper bound. 
In $d>5$ dimensions we have found that for large $T$ and $\tau$,
\begin{eqnarray}
T/2 \,&=& \int_{r_0}^{r(T/2)} \frac{dr'}{f_d(r')} \,\simeq\, r(T/2) - \alpha_d(r_0) - \mathcal{O}(r^{d-4}) \,,\\
T/2 \,&=& \int_0^{\tau/2}d\tau'\frac{E}{f_d(r(\tau'))}   \,\simeq\, \tau/2 +  \beta_d(r_0) - \mathcal{O}(\tau^{-\frac{d-5}{d-1}}) \,,
\end{eqnarray}
so the constants $\alpha_d(r_0)$ and $\beta_d(r_0)$ can be found through the following relations,
\begin{eqnarray}
\lb{alpha}
\alpha_d(r_0) \,&=& r_0 + \int_{r_0}^{r(\tau/2)} dr' \frac{f_d(r')-1}{f_d(r')} \,,\\
\beta_d(r_0) \,&=& \int_0^{\tau/2}d \tau' \frac{E-f_d(r(\tau'))}{f_d(r(\tau'))}  \,=\,  \int_{r_0}^{r_{max}(\tau)}d r' \frac{E-f_d(r')}{f_d(r')\sqrt{E^2-f_d(r')}} \,,
\lb{beta}
\end{eqnarray}
taking $\tau \rightarrow +\infty$, and where we used the fact that $r = r_0 + \int_{r_0}^{r}d r'$. Then we have for the constant $\sigma_d(r_0)$,
\be
&&\sigma_d(r_0) =  \alpha_d(r_0) + \beta_d(r_0) \nonumber \\
&&= r_0 +  \int_{r_0}^{r_{max}(\tau)}d r'\, \frac{E-f_d}{f_d}\left(\frac{1}{\sqrt{E^2-f_d}} - \frac{1-f_d}{E-f_d} \right) -  \int_{r_{max}(\tau)}^{r(\tau/2)} dr' \frac{1-f_d(r')}{f_d(r')}  \,,
\lb{sig}
\ee
where we have cut into two pieces the integral in (\ref{alpha}), from $r_0$ to $r_{max}(\tau)$ and $r_{max}(\tau)$ to $r(\tau/2)$, as $r_{max}(\tau)<r(\tau/2)$. The first integral in (\ref{sig}) is positive 
\footnote{Consider the function $h(x)=(y-x) - (1-x)\sqrt{y^2-x}$ such that $0<x\leq y^2<1$. Then $h(x)>g(x)=(y-x)\sqrt{1-x} - (1-x)\sqrt{y^2-x}$. Now, $g(x)=0$ reduces to a quadratic equation and the only solution is $x=0$. If $x=y^2$ then $g(x=y^2)=(y-y^2)\sqrt{1-y^2}>0$, which means that $g(x)>0$ if $0<x\leq y^2<1$ and hence $h(x)>0$. Providing $x=f_d$ and $y=E$ this proves that the first integral in (\ref{sig}) is positive. }
since $0< f_d(r) \leq E^2 < 1$ for $r_0\leq r \leq r_{max}$.  The last term in (\ref{sig}) is negligible because we are considering the limit $\tau \rightarrow +\infty$,
\be
\lim_{\tau \to +\infty}\int_{r_{max}(\tau)}^{r(\tau/2)} dr' \frac{1-f_d(r')}{f_d(r')} \simeq  \lim_{\tau \to +\infty}\int_{\tau^{2/(d-1)}}^{\,\tau} dr' \Big(\frac{r_s}{r'}\Big)^{d-3} = 0\,.
\ee
   As a consequence, one finds  that $\sigma_d(r_0)$ is positive,
\be
\sigma_d(r_0) > r_0 > 0\, ,
\lb{ineqsigma}
\ee
and therefore we obtain the lower bound
\be
V_{Sch^{d}}(\tau, r_0)>V_{M^{d}}(\tau)\, , \qquad d>5
\ee

\bigskip

To sum up, the volume of a large diamond in Schwarzschild spacetime is always larger that of in Minkowski spacetime, in any dimension $d\geq 4$ and independently of the position $r_0$ of the center of the generalized diamond,
\be
V_{Sch^{d}}(\tau, r_0)>V_{M^{d}}(\tau)\, .
\lb{ineqvsch}
\ee
 This inequality is actually the same as the lower bound of (\ref{MVM}) that we have found previously and is valid only for large causal diamonds. We notice a major difference between the inequality in terms of $T$ and that of $\tau$. Regarding the upper bound, we notice that in the presence of horizon the red-shift is unbounded inside the diamond. Therefore, the analog of the upper bound of (\ref{MVM}) in this case would be  trivial.

\subsection{Comparison with diamond with shifted center in Minkowski spacetime}
We can make our inequality (\ref{ineqvsch}) even stronger by comparing  $V_{Sch^{d}}(\tau, r_0)$ with the volume of a diamond with shifted center in Minkowski spacetime. 
The volume of a ``shifted diamond'' in $d$-dimensional Minkowski spacetime is given by 
\be
V_{M^{d}}(\tau, r_0) = \frac{2\Omega_{d-2}}{d(d-1)}\left((\tau/2+r_0)^d-2r_0^d\right) \simeq V_{M^d}(\tau)\left(1 + 2d\frac{r_0}{\tau} +  \mathcal{O}(\tau^{-2})\right) > V_{M^d}(\tau) \,,
\ee
for sufficiently large diamond, $\tau>2r_0$. Knowing from (\ref{ineqsigma}) that $\sigma_d(r_0) > r_0$ we get a stronger inequality than (\ref{ineqvsch}) on the volume of a large diamond in Schwarzschild spacetime,
\be
V_{Sch^{d}}(\tau, r_0) > V_{M^{d}}(\tau, r_0) > V_{M^d}(\tau) \,,
\lb{ineqVr0}
\ee
for $d>5$, providing $r_0>r_s$. This inequality is also valid for $d=4$ and $d=5$ as can be seen from (\ref{Vschshift4}) and (\ref{Vschshift5}).

Inequality (\ref{ineqsigma}) reveals itself quite useful. Indeed, combining (\ref{alpha})  and (\ref{beta}) we have for $d>5$,
\be
r(\tau/2) \simeq \tau/2 + \sigma_d(r_0) - \mathcal{O}(\tau^{-\frac{d-5}{d-1}}) \,,
\ee
then using once again $\sigma_d(r_0) > r_0$ we find that 
\be
r(\tau/2) > \tau/2 + r_0 \,, \qquad d>5\,.
\ee
This lower bound on $r(\tau/2)$ is actually universal for $d\ge4$ because of the presence in the function $T(\tau)$ of a power term and a logarithm term, in $d=4$ and $d=5$ dimensions respectively, ensuring that  
\be
r(\tau/2) > \tau/2 + r_0 \,, \qquad d\ge 4\,.
\lb{ineqr-r0}
\ee
The area is given by
\be
A_{Sch^{d}}(\tau, r_0)=\Omega_{d-2} r^{d-2}(\tau/2)\, , 
\ee
and one can infer from (\ref{ineqr-r0}) the following inequality on the area of a diamond in Schwarzschild spacetime,
\be
A_{Sch^{d}}(\tau, r_0) > A_{M^{d}}(\tau, r_0) > A_{M^{d}}(\tau) \,,
\lb{ineqAr0}
\ee
where $A_{M^{d}}(\tau, r_0)=\Omega_{d-2}(\tau/2+r_0)^{d-2}$ and $A_{M^{d}}(\tau)=\Omega_{d-2}(\tau/2)^{d-2}$ is the area in $d$-dimensional Minkowski spacetime of the ``shifted diamond'' and of the usual causal diamond, respectively. We emphasize that (\ref{ineqVr0}) and (\ref{ineqAr0}) are valid for all dimensions $d\ge4$.
 
\section{Applications}
\setcounter{equation}0
There might be a number of applications of our results. Below we consider some of them.

\subsection{Comparison theorems for entanglement entropy}
In a rather general context with each co-dimension surface $\Sigma$ in a static spacetime one can associate a reduced matrix and respectively an entropy.
This entropy is called entanglement entropy. It can be considered as a measure of correlations in a quantum system across the surface.
Among all possible co-dimension two surfaces there exist certain surfaces which can be associated with a causal diamond. This surface appears as intersection of future light cone of an event $q$ and past line cone of an event $p$ (suppose that the invariant time interval between these two events $p$ and $q$ is $\tau$).  
Not every surface can be associated with a diamond. In Minkowski spacetime only spheres are such surfaces. Suppose that $\Sigma$ is a surface associated with a diamond of duration $\tau$. The pure quantum state is associated with a spacelike hypersurface $\cal{H}$ which crosses the diamond at $\Sigma$. Then a nice feature of such a surface $\Sigma$ is that
any deformation of $\cal{H}$ for which $\Sigma$ still lies in $\cal{H}$ does not change the entanglement entropy of $\Sigma$. This is a simple consequence of the causality and that the number of degrees of freedom inside  of $\Sigma$ remains inside the causal diamond in question. The leading (UV divergent) term in the entanglement entropy is proportional to the area of $\Sigma$. Then, using our analysis above in this paper and  the inequalities  for the area, we may formulate  certain {\it comparison theorems for entanglement entropy}.  
Notice that a different comparison theorem, when the spacetime is fixed but the surface is supposed to vary, was formulated in \cite{Astaneh:2014uba}. 
For simplicity we shall consider the spherically symmetric diamonds  considered in section 3 so that we can use  the inequality (\ref{AM}) for the area. The appropriate generalizations, if desired,  can always be made.

Consider a surface $\Sigma$ associated with a causal diamond of duration $\tau$, in a curved spacetime $\cal{M}$ and in Minkowski spacetime $M$.  
For the area of this surface we have the inequality (\ref{AM}). Reformulating this inequality in terms of entanglement entropy we 
state that the entanglement entropy associated with $\Sigma$ in these two spacetimes is related by the inequalities
\be
S_\Sigma (M,\tau) \leq S_\Sigma({\cal{M}},\tau)\leq (1+z_c)^{d-2}\, S_\Sigma (M,\tau)\, ,
\lb{a1}
\ee
where $z_c$ is the maximal red-shift inside the causal diamond in the curved spacetime $\cal{M}$. The lower bound is supposed to be valid for large $\tau$ while the upper bound
for any value of duration $\tau$.

Moreover, using the inequality (\ref{i11}) mentioned in the Introduction we may formulate a comparison theorem for the entropy in de Sitter spacetime and in Minkowski spacetime.
Indeed, for the entropy associated to a diamond of same duration $\tau$ in these two spaces we have that
\be
S_\Sigma(dS,\Lambda>0,\tau)\leq S_\Sigma (M, \Lambda=0,\tau)\, .
\lb{a2}
\ee
Then, using our result (\ref{ineqAds}) we  formulate the theorem comparing the entropy in a spacetime with positive cosmological constant to that of in de Sitter spacetime, 
\be
S({\cal{M}}_\Lambda,\tau)\leq S_{dS}(\Lambda,\tau (1+z_c))\, ,
\lb{a3}
\ee
where ${\cal{M}}_\Lambda$ is a solution to Einstein equations with positive cosmological constant $\Lambda$, $z_c$ is the red-shift at the center of the diamond.
Assuming that the entanglement entropy can still be defined even if the spacetime is not static and
the leading term is still proportional to the area we may  conjecture that inequality (\ref{a3}) would still be valid even if the spacetime is not static, similarly to relation 
(\ref{i2}).

\subsection{Scattering amplitudes in asymptotically flat spacetime}
\label{scatter}
Our next remark on the possible application/generalization of our results is less concrete. It is more a suggestion for a direction in which our approach could be generalized
in the quantum field theory.
It is based on the fact that the infinite duration $\tau$ causal diamond is a natural domain for the
formulation of the scattering problem of massless particles, such as gravitons.  Therefore, we anticipate that the inequalities similar to those we have discussed in this paper should be valid for the scattering amplitudes of massless particles. More concretely, we suggest that there should exist certain comparison theorems for the 
elements of the $S$-matrix in a curved asymptotically flat spacetime and in Minkowski spacetime. We anticipate that, just like we had it for the geometric characteristics of the causal diamonds, the inequalities for the scattering amplitudes should involve the maximal red-shift in the spacetime (we consider the case of spacetime without horizons so that the red-shift takes some maximal finite value inside the infinitely large causal diamond).  It would be interesting to verify our prediction in a concrete example such as the scattering of gravitons in a weakly curved, asymptotically flat,  background.

\subsection{Causal Set Theory}

Previous work on the volumes of causal diamonds has been applied 
to Causal Set theory \cite{causalsets} and we anticipate that the  results of our
present paper will find further applications in that area. 
A causal set is a poset or partially ordered set with a transitive
relation $\prec$ such that 
$x\prec y\,,  y \prec z \, \Rightarrow  x \prec z$  which is  
 typically taken as irreflexive (i.e. such that  $ x \not\prec x $).        
For a time oriented spacetime $x\prec y$ is  the chronology 
relation $ y  \in I^+(p)$  and the Alexandrov open set or causal
diamond  $I^+(x) \cup I^-(z)$  corresponds to
the interval $(x,z) = \{y| x \prec y \prec z \}$. Causal sets 
are considerably more general than  smooth spacetime manifolds, 
typically being discrete and  the properties of causal diamonds
 may be used  to define the analogues of the 
standard geometrical objects of spacetimes which are believed to emerge
in a  continuum limit in which the causal set acquires a manifold structure
\cite{Myrheim:1978ce,papers}.
              
\subsection{Tasks and Supertasks}

Finally we would like to mention a more speculative 
potential application of the results of this paper.
Since this subject is not widely known, we provide here a brief
introduction. 

We are referring to  discussions
of the extent to which in principle the fundamental laws of physics 
impose restrictions on our ability to carry out computational
tasks.  Another relevant discussion  is
whether the limitations  imposed by the fact  that certain tasks may be not 
completed in polynomial time may be evaded by recourse to exotic 
spacetime structures such as black holes, Cauchy horizons,  or more speculatively  
closed timelike curves or wormholes. There is an extensive literature on such questions,
much of it more philosophical than physically informed, 
and we shall not attempt here to review it.
We shall simply  focus on some aspects directly connected with
the results of the present paper.
For example, the phenomenon of time 
dilation raises the interesting question of whether,
in principle, carrying out some task, for example a computation,
may be speeded up by sending an apparatus such as a computer along
a timelike curve $\gamma_e(\tau)$ of total proper time duration
$\tau_e$, while the observer  interested in the rapid outcome of the task 
moves on a timelike curve $\gamma_o(\tau)$ of proper time duration
$\tau_o$. The initial points of $\gamma_e(\tau) $ and $\gamma_o(\tau) $ 
thus coincide, $\gamma_e(0)=\gamma_o(0) =p$ 
but the endpoints $q_e=\gamma_o(\tau)$ and $q_0=\gamma_e(\tau)$
need not. However we do require that the computer is in the past
of the observer  $q_e   \in I^-(q_0)$ 
when the task is complete so that the results
can be beamed back to the home station. 
One might also demand that all of $\gamma_e(\tau) \in I^-(\gamma_o(\tau))$
so that continuous signaling of intermediate results is possible.
 
The first printed discussion of this possibility appears
to be that of McCrea \cite{McCrea}, in connection with the twin paradox.
In this case $p=q$. McCrea argued in favour of a general
``Principle of Impotence'' stating that if $\gamma_o(\tau)$
is a geodesic in free fall in Minkowski spacetime and $\gamma_e(\tau)$
undergoes acceleration, then $\tau_o > \tau_e$ otherwise 
one would  be able to circumvent the second law of thermodynamics.

One may require the same property if spacetime is curved.
Of course provided
 $\gamma_o(\tau)$ and $\gamma_e(\tau)$ are in a sufficiently
small neighborhood, it will automatically be true 
as we have demonstrated explicitly above 
in the spherically symmetric case.

The next development, apparently independently of McCrea, 
occurred when Jarett and Cover  \cite{Jarett} applied
Shannon's theorem, giving a  
lower bound on the capacity $C$ in bits per
unit time of a single
communication channel  of bandwidth $W$
capable of transmitting reliably a signal of power $S$
in the presence  of noise power
$N$. This is
\be
C \le  W \ln ( 1 + \frac{S}{ N} )\,. 
\ee
They argued for an asymmetry between  the two systems.
Roughly speaking, to transmit the same amount of information, measured in total number of bits, the world line of longer total proper time duration requires a smaller energy expenditure.
Thus, in the classic twin paradox setting, the energy cost of transmission
from  the accelerated twin is larger. 

More precisely, they argued as follows. 
 Suppose we have a stretch of signal containing $I$ bits of information
sent by an observer traveling with the computer  $e$
of duration $T_e$ and received by $o$ in a time judged by to be
$T_o$. 
Then signal power received is $S_o=(T_o/T_e) ^2 S_e$. Thus
\be
I/T_o < W_o \ln ( 1+ \frac{(T_o/T_e) ^2 S_e }{ N} )  \,.
\ee  
       
Apparently also unaware of McCrea's work, 
Pitowsky, 
Malament \cite{Malament1}  
and Hogarth \cite{Hogarth1}\cite{Hogarth2} \cite{Earman}  
especially carried the discussion to the extreme
asking whether one might carry out ``supertasks'' in this way. For example
could one check Goldbach's conjecture, that every even integer greater than
2 can be expressed as  the sum of two primes, by sending a computer
along a timelike path, such that $\tau_e= \infty$, which is
entirely contained in the past $I^-(p)$
of the end point of a curve whose 
duration $\tau_o < \infty$? Malament and Hogarth gave
two classes of physically interesting  spacetimes for which such curves exist
(see also \cite{Barrow}).
The first class
includes the Reissner-Nordstrom spacetimes. The curve $\gamma_e(\tau)$
remains outside the event horizon at $r=r_+$, while the curve $\gamma_o(\tau)$
free falls through the horizon at $r=r_+$ and then passes through the 
Cauchy horizon at $r=r_-$. The second class contains the
 universal covering space of anti-de Sitter spacetime (adS). The curve
$\gamma_o(\tau)$ is taken to be a geodesic in the bulk while
$\gamma_e(\tau)$ is taken to be a timelike curve which ends on the timelike
 conformal boundary. 
In fact it is easy to see, using the methods of \cite{Geroch}, 
that adS shares with
Minkowski spacetime the property that any timelike curve reaching the 
conformal boundary has
\be
\int _{\gamma_e(\tau)} |a| d\tau = \infty \,,
\ee
where $|a|$ is the magnitude of the 4-acceleration.
Thus if the acceleration is bounded, then $\tau_e =\infty$. The difference
between adS  and  Minkowski spacetime  is that in the latter case,
a curve reaching future null infinity ${\cal I } ^+$ is only visible 
from a timelike geodesic $\gamma_o(\tau)$ in the bulk of infinite total
proper time duration, $\tau_o =\infty$. Note that for a physically
feasible curve with conventional fuel,  the fuel
consumption is bounded below by the inequality
\be
{ m_f \over m_i} < \exp( - \int |a| d \tau)\,,
\ee
where $m_i$ and $m_f$ are the initial and final masses respectively.
Thus purely on those grounds the proposed scenario is clearly
not physically feasible. Another obvious limitation
on  accelerating  computers  is that they will
experience  thermal fluctuations due to Hawking-Unruh effects
which will degrade their performances \cite{Narnhofer,Deser1,Deser2,Deser3}.    

For the reason stated above,  examples  involving  world lines of infinite
duration seem rather contrived. It is more  
reasonable from a physical point
of view to consider a computing running for a finite proper time.  
Such an operation entails
sending instructions  from some sort of central processor
and subsequently receiving the outcome of those instructions.
The largest domain of spacetime available to the central processor for this  
processing is the causal diamond whose endpoints are the beginning and end
of the world line under consideration. Quantitatively it 
seems reasonable to take the volume $V(\tau)$ as a measure of the
largest amount of information that can be processed in a  total time
$\tau$. The natural comparison is with the volume $V_M(\tau)$ 
of a causal diamond in Minkowski spacetime (cf. \cite{Biamonte}).   

The results of our previous papers 
\cite{Gibbons:2007nm,Gibbons:2007fd,Solodukhin:2008qx} show how the local 
Ricci curvature and hence the local energy momentum tensor
affects the running of our hypothetical computer.
In four spacetime dimensions, for small causal diamonds filled with a  
 perfect fluid
we found (cf. \cite{Gibbons:2007nm} eq.~(41))   that
\be
V(\tau) = V_M(\tau)  
 \Big( 1 +\frac{4€\pi G}{45} 
(\rho +6P)\tau ^2  +\dots  \Big) \,,
\ee  
and so as long as $ P> -\rho/6$ the performance
of our hypothetical computer will be improved by being immersed 
in a self-gravitating medium. 

For a static spacetime, the results of our present
paper illustrate how  the redshift at the location   of our hypothetical 
computer affects its operation. Specifically (\ref{spatV}) 
shows that lowering our computer into a region  of low
gravitational potential may improve the functioning of 
our hypothetical computer, but the amount of improvement is
bounded by the fourth power of the redshift factor.
It might well prove rewarding to explore the possible relevance to
the wider considerations reported in \cite{Nemeti}.
It is even possible that one could envisage a computer
which operates by carrying out scattering experiments.
The ideas  of  section \ref{scatter} concerning  large causal diamonds 
might then be relevant.

\section{Summary} In this paper we have formulated a number of
statements in the form of inequalities that compare the geometric
characteristics of the causal diamonds of fixed duration $\tau$ in
spacetime curved by positive energy matter and in Minkowski spacetime.
These statements are analogs of the well-known Bishop's inequalities
that exist in the Euclidean geometry.  As we have observed, in the
absence of horizons,  the curved spacetime has larger volume than in
Minkowski spacetime. This is also true for the spatial volume and the
area. On the other hand, the ratio of two volumes is limited from
above by the maximal  red-shift factor inside the diamond. These
statements are generalized for spacetime with (positive or negative)
cosmological constant. The comparison in this case is made with the
maximally symmetric (anti-) de Sitter spacetime. The comparison
between the curved space with cosmological constant $\Lambda$ and
Minkowski spacetime also can be done.  For positive $\Lambda$ the
volume is always less than the volume in Minkowski spacetime while
for negative $\Lambda$ the volume is larger than that of in Minkowski
spacetime.  We have suggested a generalization of these statements for
spacetimes with horizons. More concretely we have considered the
Schwarzschild $d$-dimensional metric and introduced a generalized
diamond with shifted center. In the limit of large duration of the
generalized diamond we have formulated same inequalities as in the
case without horizons.

We have suggested some applications of these results. For instance we
have formulated some comparisons theorems for the entanglement entropy
associated with a diamond of fixed duration $\tau$. These theorems can
be possibly verified independently using for instance the holographic
methods.   We also anticipate that similar comparisons can be done for
the scattering amplitudes of massless particles such as gravitons. It
would be interesting to make these predictions more concrete. Similarly
it would be desirable to tighten up our
ideas about causal diamonds and the limits of computation.  Clearly,
more work needs to be done in the future.

\section*{Acknowledgments} 
C.B. is supported by the CNRS.

\end{document}